\documentclass[aps,pra,reprint,superscriptaddress,longbibliography]{revtex4-1}

\usepackage{amsmath}
\usepackage{amsfonts}
\usepackage{amssymb}
\usepackage{ulem}
\usepackage{graphicx}
\usepackage{color}
\usepackage[dvipsnames]{xcolor}
\usepackage[colorlinks=true,citecolor=blue,linkcolor=blue,urlcolor=blue]{hyperref}
\usepackage{times}
\usepackage{amstext}
\usepackage{physics}
\usepackage[T1]{fontenc}

\newcommand{\sma}{\sigma_{+}}
\newcommand{\sme}{\sigma_{-}}

\newcommand{\blue}[1]{{#1}}

\begin{document}

\title{Estimating the degree of non-Markovianity using machine learning}

\author{Felipe F. Fanchini}
\email{fanchini@fc.unesp.br}
\affiliation{Faculdade de Ci\^encias, UNESP - Universidade Estadual Paulista, Bauru, SP, 17033-360, Brazil}

\author{G\"oktu\u{g} Karpat}
\affiliation{Faculty of Arts and Sciences, Department of Physics, \.{I}zmir University of Economics, \.{I}zmir, 35330, Turkey}

\author{Daniel Z. Rossatto}
\affiliation{Universidade Estadual Paulista (Unesp), Campus Experimental de Itapeva, 18409-010 Itapeva, S\~{a}o Paulo, Brazil}

\author{Ariel Norambuena}
\affiliation{Centro de Investigaci\'{o}n DAiTA Lab, Facultad de Estudios Interdisciplinarios, Universidad Mayor, Chile}

\author{Ra\'{u}l Coto}
\affiliation{Centro de Investigaci\'{o}n DAiTA Lab, Facultad de Estudios Interdisciplinarios, Universidad Mayor, Chile}

\date{\today}

\begin{abstract}
In the last years, the application of machine learning methods has become increasingly relevant in different fields of physics. One of the most significant subjects in the theory of open quantum systems is the study of the characterization of non-Markovian memory effects that emerge dynamically throughout the time evolution of open systems as they interact with their surrounding environment. Here we consider two well-established quantifiers of the degree of memory effects, namely, the trace distance and the entanglement-based measures of non-Markovianity. We demonstrate that using machine learning techniques, in particular, support vector machine algorithms, it is possible to estimate the degree of non-Markovianity in two paradigmatic open system models with high precision. Our approach can be experimentally feasible to estimate the degree of non-Markovianity, since it requires a single or at most two rounds of state tomography.
\end{abstract}

\maketitle

\section{Introduction}

Artificial intelligence is a wide research field that simulates human intelligence using certain machines that are programmed to perform human-like skills. Presently, the study of artificial intelligence encompasses several branches, such as machine learning (ML), reinforcement learning and deep learning, the former being the most prominent one. ML has recently become a crucial tool for extracting useful information from the very rapidly increasing rate of available data, and is now widely used in numerous research areas including computer science, medicine, chemistry, biology and physics~\cite{jordan255}. \blue{In particular, supervised ML is an approach where a training data set is introduced to the ML algorithm so that it can learn to yield the desired outputs. When provided with a training data set that includes certain inputs and their correct outputs, the model can learn over time to provide accurate estimations of the output also for the inputs that have not been used in the training process~\cite{scikit-learn}. In case of a classification problem, these outputs can correspond to different classes such as Markovian and non-Markovian quantum dynamics. On the other hand, a regression problem deals with outputs that corresponds to real values such as the outcomes of a measurement, where the aim is to make reliable projections about the desired measurement outcome.}
%On the other hand, the possible outputs might become, for instance, the outcomes of a measurement for a considered regression problem, where the aim is to make reliable projections about the desired measurement outcome.} 
Recently, ML has had a remarkable impact in physics~\cite{Dunjko_2018, Mehta_2019, Carleo_2019}, for instance, in the fields of condensed matter physics~\cite{Ghiringhelli2015,Torlai2016,Carleo2017}, quantum phase transitions~\cite{Carrasquilla_2017, Ponte_2017, Liu_2019, Canabarro_2019}, and quantum information science~\cite{Torlai2018,Canabarro2019,Raban2020}.

Unlike the ideal-isolated quantum systems that evolve unitarily in time, realistic quantum systems are in general open to interaction with an environment, which gives rise to non-unitary dynamics resulting in loss of coherence~\cite{BreuerPet,Rivas2012}. Understanding the physics of open quantum systems is of both fundamental and practical interest since the development of quantum technologies relies on the presence of precious quantum resources such as coherence~\cite{Baumgratz2014,Streltsov2017}. One of the principal concepts in the study of open quantum systems is the dynamical memory effects which might arise throughout the time evolution of the open system, and define non-Markovian dynamics~\cite{BreuerPet}. Although, under special circumstances, the evolution of open systems can be treated under Markovian approximation ignoring the memory effects, non-Markovian behavior cannot be overlooked in many realistic settings. In fact, the theory of non-Markovianity in the dynamics of open quantum systems has been widely explored in recent literature from various perspectives~\cite{Breuer2016,Li2018,Li2020x,Fanchini2013,Addis2016}, and numerous means of quantifying it have been proposed~\cite{Rivas2014}. In addition, the detection of memory effects in the open system dynamics has also been experimentally achieved~\cite{Liu2011,Fanchini2014,Haseli2014,Li2020}. More recently, ML methods have been started to be employed to study non-Markovian quantum processes~\cite{Banchi2018,Shrapnel2018,Luchnikov2020,2004.11038}. 

In this work, we present a computational approach based on ML techniques to determine the degree of non-Markovianity in the dynamics of open systems. The proposed approach requires prior knowledge only about the type of dominant decoherence process that our system of interest undergoes. In other words, we will assume that we know the underlying dynamical process, but we have no information about the characteristic model parameters defining the Markovian or non-Markovian nature of this process. Here, we consider two distinct well-established quantifiers of memory effects for our analysis, namely, the trace distance~\cite{Breuer2009} and the entanglement-based measures~\cite{Rivas2010}. Although capturing the signatures of non-Markovian behavior has been possible in some experiments in recent literature~\cite{Li2020}, accurate determination of the degree of non-Markovianity remains challenging for a wide variety of experimental setups, since in general it would require a large number of rounds of quantum state tomography to be successfully performed on the open system. Moreover, depending on the considered measure in an experiment, one would need to deal with the time evolution of a pair of different initial states or even introduce an ancillary system that needs to be protected from the destructive environmental effects. Consequently, the main motivation of our study is to simplify the experimental determination of the non-Markovianity quantifiers with the help of a ML algorithm. In particular, we show that a support vector machines (SVM) based model precisely assesses the degree of non-Markovianity of two paradigmatic quantum processes, i.e., phase damping (PD) and amplitude damping (AD), with only a single or at most two rounds of quantum state tomography. At the same time, our results provide a proof of principle that the non-Markovianity quantifiers can be precisely estimated with the assistance of ML techniques.

This manuscript is organized as follows. In Sec.~\ref{sec2}, we introduce the quantifiers of non-Markovianity considered in our work. Sec.~\ref{sec3} includes the open system models we consider in our analysis. In Sec.~\ref{sec4}, we briefly review the ML model we use in our analysis. We present our main results in Sec.~\ref{sec5} and we conclude in Sec.~\ref{sec6}. { Details of the SVM-based ML approach are discussed in the appendix.}

\section{Quantifying Non-Markovianity} \label{sec2}

{

\blue{In this section, we intend to elaborate on the characterization and quantification of non-Markovianity in the dynamics of open quantum systems. Before going into the details of the non-Markovianity measures and the notion of memory effects that we consider in our analysis, let us first discuss the fundamental and practical relevance of the non-Markovianity measures in quantifying the degree of memory effects in the dynamics of open systems.

To start with, memory effects are known to play an important role in certain significant quantum information protocols. For instance, considering an optical physical setup, it has been shown that in the case of mixed state teleportation under decoherence, increasing the amount of memory in the open system dynamics enhances the fidelity of the protocol, even allowing for perfect teleportation~\cite{Laine2014}. In a similar setting, it has also been experimentally demonstrated that superdense coding under noise can be efficiently performed due to the emergence of memory effects in the dynamics ~\cite{Liu2016}, where the superdense coding capacity can actually be expressed as a function of the trace distance measure. Moreover, it has been shown that, in a realistic open system scenario, the amount of memory in the dynamics directly controls the lower bound of the entropic uncertainty relation~\cite{Karpat2015}, which is in turn related to applications such as witnessing entanglement and cryptographic security~\cite{Berta2010}. In addition, utilizing Landauer’s principle, it has been argued that the degree of memory effects determine the amount of work extraction by erasure under decoherence~\cite{Bylicka2016}. We also mention that a rather general framework has been introduced in Ref.~\cite{Bylicka2014}, where greater values of non-Markovianity has been shown to induce larger revivals of classical and quantum capacities which would potentially improve error correction schemes. Besides, it has been very recently demonstrated that the emergence of spontaneous quantum synchronization between a pair of two-level systems (which has consequences such as the creation of robust quantum correlations between the pair~\cite{giorgi2012}), can be delayed and even completely prevented as a consequence of the increasing degree of non-Markovianity in the dynamics~\cite{Karpat2020}. Finally, we emphasize that non-Markovianity in the quantum regime is multifaceted phenomenon and different measures can be relevant in different physical problems.}

Despite the established definition of non-Markovianity in classical settings, non-Markovianity in the quantum regime is a rather delicate phenomenon~\cite{Vacchini2011}. Traditionally, a prototypical Markovian quantum process is defined based on the Lindblad type master equation, which gives rise to a semigroup of completely positive quantum dynamical maps~\cite{Lindblad1976,gorini1976}. A more general class of quantum processes satisfies the property of completely positive divisibility (CP-divisibility) in connection with the non-negativity of the decay rates in time dependent Lindblad master equations~\cite{Breuer2012}. Let us assume that we have a dynamical quantum process $\Lambda$, i.e., a completely positive trace preserving (CPTP) map, describing the time evolution of a quantum system. In recent literature, Markovian quantum dynamical maps are typically considered to be the ones which obey the decomposition law $\Lambda(t,0)=\Lambda(t,s)\Lambda(s,0)$ where, in addition to $\Lambda(t,0)$ and $\Lambda(s,0)$, the transformation $\Lambda(t,s)$ is also a CPTP map for all $s\leq t$. Such maps are known as CP-divisible transformations and are said to imply a memoryless evolution for the open system. Therefore, based on the violation of the decomposition relation (or equivalently the degree of violation of the CP-divisibility property), it becomes possible to define quantifiers to measure the degree of non-Markovianity in  open system dynamics.

At this point, it is important to emphasize that most of the non-Markovianity quantifiers in the literature are actually witnesses for the breakdown of CP-divisibility rather than strict measures~\cite{Rivas2014}. In other words, even though these quantifiers consistently vanish when CP-divisibility property is satisfied, they are not always guaranteed to capture its violation. However, we should recall that some of these non-Markovianity witnesses might still be considered as non-Markovianity measures (or measures for the degree of memory effects in the dynamics) on their own the right, since they can be used for quantifying the backflow of information from the environment to the open system, which by itself can be used as a basis for the definition of non-Markovian dynamics in the quantum regime~\cite{Breuer2016}. This approach is also intuitive because in this way the future states of an open system can depend on its past states, due to the flow of information from the environment back to the open system during the time evolution~\cite{Breuer2009,Fanchini2014}.

Having briefly elaborated on what we mean by memory effects, we are in a position to discuss the non-Markovianity measures that we consider in our study. Let us first introduce the trace distance measure which is constructed upon the distinguishability of an arbitrary pair of quantum states represented by the density operators $\rho_1$ and $\rho_2$. Trace distance between these two states can be written as $D(\rho_1,\rho_2)=\frac{1}{2} \rm{Tr}|\rho_1-\rho_2|$, with $|A|=\sqrt{A^\dagger A}$. Since a temporary increase of distinguishability, measured with the trace distance, throughout the open system dynamics can be interpreted as a backflow of information from the environment to the open system, signatures of memory effects are signaled when $dD/dt>0$. On the other hand, if the trace distance monotonically decreases or remains constant during the dynamics, that is $dD/dt \leq 0$, then it means that the dynamics has no memory and thus it is Markovian. Therefore, the degree of non-Markovianity can be measured by~\cite{Breuer2009}
\begin{equation}\label{NBreuer}
\mathcal{N}_{D}=\max_{\rho_1(0),\rho_2(0)}\int_{(dD(t)/dt)>0}\frac{dD(t)}{dt}dt
\end{equation}
where the optimization is performed over all possible pairs of initial states $\rho_1(0)$ and $\rho_2(0)$. As it has been suggested in the recent literature~\cite{Addis2014}, in our calculations we assume that the optimal initial states are orthogonal~\cite{Wismannn2012} and given by the eigenstates of the Pauli operator along $x$ direction.  We recall that due to the fact that the trace distance is contractive under CPTP transformations, distinguishability between $\rho_1$ and $\rho_2$ monotonically decreases for all CP-divisible dynamical maps at all times. However, as mentioned earlier, non-Markovianity based on the trace distance measure is not equivalent to the breakdown of CP-divisibility property.

The second measure that we use in our study is based on the entanglement dynamics of a bipartite quantum state, given by our system of interest and an ancilla that is isolated from the effects of the environment. Aside from the interpretation of the information flow using distinguishability, this approach is linked to the information dynamics between the open quantum system and its environment through entropic quantities~\cite{Rivas2010,Fanchini2014}. Specifically, let us introduce an ancilla system $A$, which has the same dimension as the principal open system $B$. Considering that the subsystem $B$ undergoes decoherence and the ancilla $A$ trivially evolves, a monotonic decrease in entanglement of the bipartite system $AB$ implies that the dynamics is Markovian. However, any temporary increase in entanglement throughout the time evolution can be used to capture the memory effects in the open system dynamics. Thus, non-Markovianity can be quantified with
\begin{eqnarray}\label{NEntanglement}
\mathcal{N}_E&=&\max_{\rho_{AB}(0)}\int_{(dE(t)/dt)>0}\frac{dE(t)}{dt}dt
\end{eqnarray}
where $E$ denotes an entanglement measure and the optimization is carried out over all initial states of the bipartite system $\rho_{AB}(0)$. Since it has been demonstrated for a single qubit open system and an ancilla that the optimal value of the measure is attained for Bell states~\cite{Neto2016}, we calculate it considering that the initial bipartite system $AB$ is in one of the Bell states. In fact, any entanglement measure can be used to evaluate this measure. Here we choose to focus on the concurrence~\cite{Wootters1997}. We should also finally note that as entanglement measures are monotones under local CPTP maps, the entanglement based non-Markovianity measure vanishes for all CP-divisible processes, similar to the trace distance measure.}

\section{Open Quantum System Models} \label{sec3}

We now introduce the paradigmatic open quantum system models that we consider to study how well one can determine the degree of non-Markoviantity using ML techniques.

\subsection{Phase Damping}

Let us first consider a two-level quantum system (qubit) undergoing decoherence induced by colored dephasing noise as introduced in Ref.~\cite{daffer04}. Suppose that the time-evolution of the qubit is described by a master equation of the form
\begin{equation} \label{mem}
\dot{\rho}=K\mathcal{L}\rho,
\end{equation}
where $\mathcal{L}$ is a Lindblad superoperator and $\rho$ denotes the density operator of our system of interest. Here, the time-dependent integral operator $K$ acts on the open system as $K\phi=\int_0^t k(t-t')\phi(t')dt'$ with $k(t-t')$ being a kernel function governing the type of memory in the environment. A master equation of the form given in Eq.~(\ref{mem}) can arise, for instance, when one considers a time-dependent Hamiltonian
\begin{equation}
H(t)=\hbar\sum_{k=1}^3\Gamma_k(t)\sigma_k,
\end{equation}
where $\Gamma_k(t)$ are independent random variables possessing the statistics of a random telegraph signal, and $\sigma_k$ are the Pauli matrices in $x, y$ and $z$ directions. The random variables can be expressed as $\Gamma_k(t)=\alpha_k n_k(t)$, where each $n_k(t)$ has a Poisson distribution with a mean equal to $t/2\tau_k$ and $\alpha_k$ is a coin-flip random variable with the possible values $\pm \alpha_k$. While $\alpha_k$ describe the coupling of the open system to the random noise, the flipping rates are given by $1/\tau_k$.

To obtain a solution for the density operator $\rho$ of the open system qubit, one can directly use the von Neumann equation given by $\dot{\rho}=-(i/\hbar)[H,\rho]$, then it reads
\begin{equation}
\rho(t)=\rho(0)-i \int_0^t\sum_k \Gamma_k(s)[\sigma_k,\rho(s)]ds. \label{isol}
\end{equation}
Substituting Eq. (\ref{isol}) back into the von Neumann equation and evaluating the stochastic average, one gets
\begin{equation}
\dot{\rho}(t)=-\int_0^t\sum_k e^{-(t-t')/\tau_k}\alpha_k^2 [\sigma_k,[\sigma_k,\rho(t')]]dt', \label{sol}
\end{equation}
using the correlation functions of the random telegraph signals $\langle\Gamma_j(t)\Gamma_k(t')\rangle=\alpha_k^2\exp(-|t-t'|/\tau_k)\delta_{jk}$, which define the memory kernel. In Ref.~\cite{daffer04}, it has also been shown that under the condition that the noise acts only in a single direction, i.e., when two of the $\alpha_k$ vanish, the dynamics generated by Eq. (\ref{sol}) is completely positive. In fact, if $\alpha_3=1$ and $\alpha_1=\alpha_2=0$, then the open system undergoes decoherence induced by a colored dephasing noise. In this case, the Kraus operators describing the dynamics of the open system are given by
\begin{align}
M_1(\nu) &= \sqrt{[1+\Lambda(\nu)]/2}\mathbb{I}, \\
M_2(\nu) &= \sqrt{[1-\Lambda(\nu)]/2}\sigma_3,
\end{align}
where $\Lambda(\nu)=e^{-\nu}[\cos(\mu\nu)+\sin(\mu\nu)/\mu]$, $\mu=\sqrt{(4\tau)^2-1}$, $\nu=t/2\tau$ is the dimensionless time and $\mathbb{I}$ denotes the identity operator. Particularly, the dynamics of the open system can be expressed using the operator-sum representation as
\begin{equation}
\rho(\nu) = \sum_{i=1}^{2}M_{i}(\nu)\rho(0)M_{i}^{\dagger}(\nu).
\end{equation}
We note that the parameter $\tau$ controls the degree of memory effects responsible for the emergence of non-Markovianity, that is, as $\tau<1/4$ gives a Markovian time evolution, $\tau>1/4$ implies a non-Markovian dynamics for the open system, {according to both measures that we have introduced}. For further details about the physical relevance of the considered model in this part, interested readers might refer to Ref.~\cite{daffer04}.

\subsection{Amplitude Damping}

We will now consider a resonantly driven qubit under the influence of an AD channel, which is modelled as a bosonic reservoir at zero temperature~\cite{whalen2016,Haikka2010,Haikka2010-2,Shen2014,Huang2017}. The dynamics for this configuration is described by the Hamiltonian ($\hbar=1$)
\begin{align} 
H &= \omega_{0}\sma\sme + \Omega(\sma e^{-i\omega_L t} + \sme e^{i\omega_L t}) \nonumber \\
&+ \sum\nolimits_{k} \omega_{k}a_{k}^{\dag}a_{k} + \sum\nolimits_{k} (  g_{k}^{\ast}\sma a_{k}+g_{k}\sme a_{k}^{\dag}) \label{Hamiltonian1}, 
\end{align}
where $\sma =\sme^\dagger= \ket{\rm e}\bra{\rm g}$, and $\ket{\rm e}$ ($\ket{\rm g}$) corresponds to the excited (ground) state of the qubit with transition frequency $\omega_0$. The external driving field strength and its frequency are denoted by $\Omega$ and $\omega_L = \omega_0$, respectively, while $a_{k}^{\dagger}$ ($a_{k}$) is the creation (annihilation) operator of the $k$-th reservoir mode with frequency $\omega_{k}$. Finally $g_{k}$ is the coupling strength between the qubit and the $k$-th mode. The dissipation kernel is given by
\begin{align}
f(t)  &= \sum\nolimits_{k} \left\vert g_{k}\right\vert ^{2}e^{-i\left(  \omega_{k}-\omega_{0}\right)  t} \nonumber\\
&= \int\nolimits_{0}^{\infty}d\omega J\left(  \omega\right)  e^{-i\left(  \omega-\omega_0\right)t},\label{ft}
\end{align}
with $J\left(\omega\right)$ being the spectral density of the reservoir. Without loss of generality, we assume the qubit resonantly couples to a reservoir with a Lorentzian spectral density \cite{BreuerPet,whalen2016,Haikka2010,Haikka2010-2,Shen2014,Huang2017,Bellomo2007}
\begin{equation}
J( \omega)  =\left(  \frac{\gamma_0}{2\pi}\right)  \frac{\lambda^{2}}{\left(  \omega-\omega_0\right)  ^{2}+\lambda^{2}}, \label{spectraldensity}
\end{equation}
in which the spectral width (twice the coupling $\lambda$) is related to the correlation time of the reservoir $\tau_{B}\approx1/\lambda$, whereas $\gamma_0$ is connected to the time scale in which the state of the system changes $\tau_{R}\approx1/\gamma_0$~\cite{BreuerPet}. For this spectral density and considering no external field, the open system dynamics is essentially Markovian within the weak coupling regime, which corresponds to $\tau_{R}>2\tau_{B}$ $\left( \lambda>2\gamma_0\right) $. By contrast, the dynamics exhibits non-Markovian features within the strong coupling regime where $\lambda<2\gamma_0$ { for both of the considered measures}.

When the spectral density is Lorentzian, the interaction of the qubit with its genuine environment can be exactly modeled by an equivalent `Markovian' description, in which the qubit itself is coupled to a damped harmonic oscillator (auxiliary pseudomode described by the bosonic operators $b$ and $b^{\dagger}$), which is initially in the vacuum state. Relationship between the original environment variables and the psedomode ones is well established and the details can be found in Ref.~\cite{Garraway1997}; besides it is worth emphasizing the pseudomode is a mathematical construction and, strictly, does not exist physically. Here, the system-pseudomode dynamics, described by the density operator $\varrho_t$, is given by the following master equation in a frame rotating with the driving field frequency~\cite{whalen2016}
\begin{equation} \label{mateqeff}
	\dot{\varrho}_t = -i[\mathcal{H},\varrho_t] + {\mathcal{L}}_b \varrho_t,
\end{equation}
with
\begin{align} 
	&\mathcal{H} = \Omega(\sma +\sme) + \sqrt{\lambda \gamma_0/2}\,(\sma b + b^\dagger \sme), \label{Heff} \\
	&{\mathcal{L}}_b \varrho_t = \lambda(2b \varrho_t b^\dagger - b^\dagger b \varrho_t - \varrho_t b^\dagger b) \label{Lb}.
\end{align}
The qubit dynamics is obtained by taking the partial trace over the harmonic oscillator degrees of freedom, i.e., $\rho_t = \Tr_{b}[\varrho_t]$. We remark that, up to our best knowledge, Eq.~\eqref{mateqeff} does not have a closed-form solution for $\rho_t$ in the general case. However, when there is no external driving field, $\Omega=0$, open system dynamics of the qubit is then given by~\cite{BreuerPet,Bellomo2007}
\begin{equation}
\rho_t = \begin{pmatrix}
        \rho_\text{ee}^{0} P_t \,\,\, & \,\,\, \rho_\text{eg}^{0} \sqrt{P_t} \\
        \rho_\text{ge}^{0} \sqrt{P_t} \,\,\, & \,\,\, \rho_\text{gg}^{0}+\rho_\text{ee}^{0} (1-P_t)
    \end{pmatrix}, \label{rhotamp}
\end{equation}
where $P_t = e^{-\lambda t}[\cos(dt/2)+(\lambda/d)\sin(dt/2)]^2$ with $d=\sqrt{2\gamma_0\lambda - \lambda^2}$, and $\rho_{ij}^{0}$ denotes the initial state elements.

\section{Machine Learning} \label{sec4}

There are now myriads of learning models available in the literature~\cite{scikit-learn}, each of which are suitable for a particular problem. Since we will perform our calculations using SVM throughout this study, it is instructive to {briefly} introduce the main aspects of this computational approach. \blue{A more elaborated explanation about SVM, including a more illustrative and pedagogical example, is given in Appendix~\ref{app}.}

\subsection{Support Vector Machines}

One of the most well understood ML models is SVM~\cite{Vapnik_1995}. This model can be used for classification (SVC)~\cite{Burges_1998,Dietrich_1999,Risau_2000, Opper_2001} and regression (SVR)~\cite{Scholkopf_1998, Scholkpf_2002, Smola_2004, Drucker96}. Moreover, it has been recently extended to the quantum regime~\cite{Rebentrost_2014, Li_2015, Biamonte_2017, Havlicek_2019}. In general lines, SVC is a class of algorithms aiming to find a hyperplane that splits the dataset based on the different classes. Therefore, predicting the label of unknown data is relatively easy, since it only depends on where the data samples fall with respect to the hyperplane. The way a hyperplane can be defined is not unique, and thus, SVC sets the maximum-margin, i.e. maximizing the distance  between the hyperplane and some of the boundary training data, which are the data samples that are close to the edge of the class. These particular samples are known as support vectors (SVs). Since SVs are a subset of the training data set, this model is suitable for situations where the number of training data samples is small as compared to the dimension of the features vector. Moreover, once the model has fitted the training data set, it can be used as a decision function that predicts new samples, without holding in memory the training data set. For a non-linearly-separated data set, it is possible to define a kernel function that takes the samples to a higher dimensional space, where they are linearly separated. Although we have only provided an intuitive representation for SVC, here we give a brief mathematical description for SVR which will be our main tool in the rest of this manuscript.

SVR delivers the tools for finding a function $f(\textbf{x})$ that fits the training data set $\lbrace \textbf{x}_i,y_i \rbrace$, where $\textbf{x}_i\in\mathbb{R}^d$, and $y_i\in\mathbb{R}$ labels each sample. Note that $d$ stands for the dimension of the features vector. For illustration, we focus on a linear function $f(\textbf{x})= \textbf{w}\cdot\textbf{x} + b$, with $\textbf{w}\in\mathbb{R}^d$ and $b\in\mathbb{R}$ being fitting parameters. For $\epsilon$-SVR \cite{Vapnik_1995},  deviations of $f(\textbf{x}_i)$ from the labeled data ($y_i$) must be smaller than $\epsilon$, i.e. $\vert f(\textbf{x}_i)-y_i\vert\leq\epsilon$. Moreover, the desired function must be as flat as possible but can also include some errors. Therefore, the optimization problem can actually be stated as \cite{Vapnik_1995,Smola_2004,scikit-learn}
\begin{eqnarray}
&\mbox{minimize} \hspace*{1cm}  \frac{1}{2} \left\Vert \textbf{w}\right\Vert ^2 + C\sum_i \left( \xi_i + \xi_i^\ast \right)\label{min01} \\
&\mbox{subjected to} \hspace*{1cm} \left\lbrace \begin{array}{l}
  y_i - \textbf{w}\cdot\textbf{x}_i -b \leq \epsilon + \xi_i \\
  \textbf{w}\cdot\textbf{x}_i +b - y_i\leq \epsilon + \xi_i^\ast \\
 \xi_i,\xi_i^\ast \geq 0
\end{array} \right. \label{min02}
\end{eqnarray}
where { $||\cdot ||^2$ stands for the squared Euclidean distance,} $\xi_i,\xi_i^\ast$ are real slack variables and the condition $C>0$ sets the tolerance for deviations larger than $\epsilon$. 

{
Before we start to present our main results using the considered SVM model, we would like to first define certain terms, which are commonly used in ML studies, for the readers who might be unfamiliar with the subject. In our work, the \textit{regressor} is an algorithm which basically estimates the relationship between independent input variables and a certain output variable. While these independent variables acting as input data are known as \textit{features}, the output of the regressor is said to be the \textit{target} value. When a training data set including features and their respective target values is introduced to the ML algorithm, it attempts to find patterns in this set to create a regressor. This is known as the process of training, during which the algorithm learns from the training data set. In other words, a learning algorithm such as SVM takes the training data and produces a regressor which can in turn give reliable predictions for the output values of independent inputs. In our study, the features will be expectation values of spin observables at certain times, and the target value will be the degree of non-Markovianity of the considered open quantum system dynamics. \blue{It is important to note that the target value cannot be evaluated as a simple function of the features since the expectation values of the observables are not explicitly connected to the degree of non-Markovianity. Consequently,} in Appendix~\ref{app}, we elaborate on how the SVM based ML algorithm functions by first providing a simple illustrative example and then discussing its mathematical details.}

\section{Main Results} \label{sec5}

We commence our analysis considering what we refer to as pure PD and AD channels, where a pure channel means that no external driving field is present. For each one of these models, in order to generate a database for the training process, we calculate the time evolution of the open system and use the aforementioned measures to quantify the degree of non-Markovianity for model parameters, i.e., $\lambda$ and $\tau$. We consider a wide range of parameter values that define the two processes. In particular, in case of the AD channel, we consider the coupling parameter $\lambda/\gamma_0$ to be in the range $[0.1,3.0]$ with a step size equal to $10^{-3}$, which will enable us to generate a uniformly distributed training data with 2900 samples. On the other hand, for the PD channel, the parameter $\tau$ is varied in the range $[0.1,0.5]$ with a step size equal to $10^{-4}$, which will result in a uniformly distributed training data with 4000 samples. Hereafter, we name each sample of these databases as $\lambda_n$ and $\tau_n$. It is worth to note that we actually create two independent regressors, one for each channel, but we discuss both of them in parallel because of the identical procedure.

Next, we calculate the expectation values $\mathcal{O}_x$, $\mathcal{O}_y$, and $\mathcal{O}_z$ at a fixed time $t^*$ in the dynamics where 
\begin{equation}
\mathcal{O}_k = {\rm{Tr}}[\sigma_k\rho(t^*)],
\end{equation}
with $\sigma_k$ being the three Pauli spin operators in the $x$, $y$ and $z$ directions. We should emphasize that the expectation values for $\mathcal{O}_k$ are calculated for all $\lambda_n$ and $\tau_n$ individually at each fixed time point $t^*$. Therefore, our database now contains, for each model parameter, the expectations values $\mathcal{O}_x(t^*)$, $\mathcal{O}_y(t^*)$, and $\mathcal{O}_z(t^*)$ as the \textit{features} and the degree of non-Markovianity $\mathcal{N}$ as our \textit{target} value. We note that the experimental determination of these expectation values can be realized with a single quantum state tomography performed at each time $t^*$. \blue{We should also emphasize that, to train the regressor by providing it with a data set composing of the features and their target values, the degree of non-Markovianity is calculated numerically employing the definitions given in Eq.~(\ref{NBreuer}) and Eq.~(\ref{NEntanglement}).} To summarize, we introduce to our learner a set of features and their known respective targets. Our main objective will be to produce a regressor that will be able to determine the degree of non-Markovianity, given a pure decoherence process (without external fields), using only the information contained in the expectation values {at a fixed time}.

\begin{figure}[t]
\centering
\includegraphics[width=0.48\textwidth]{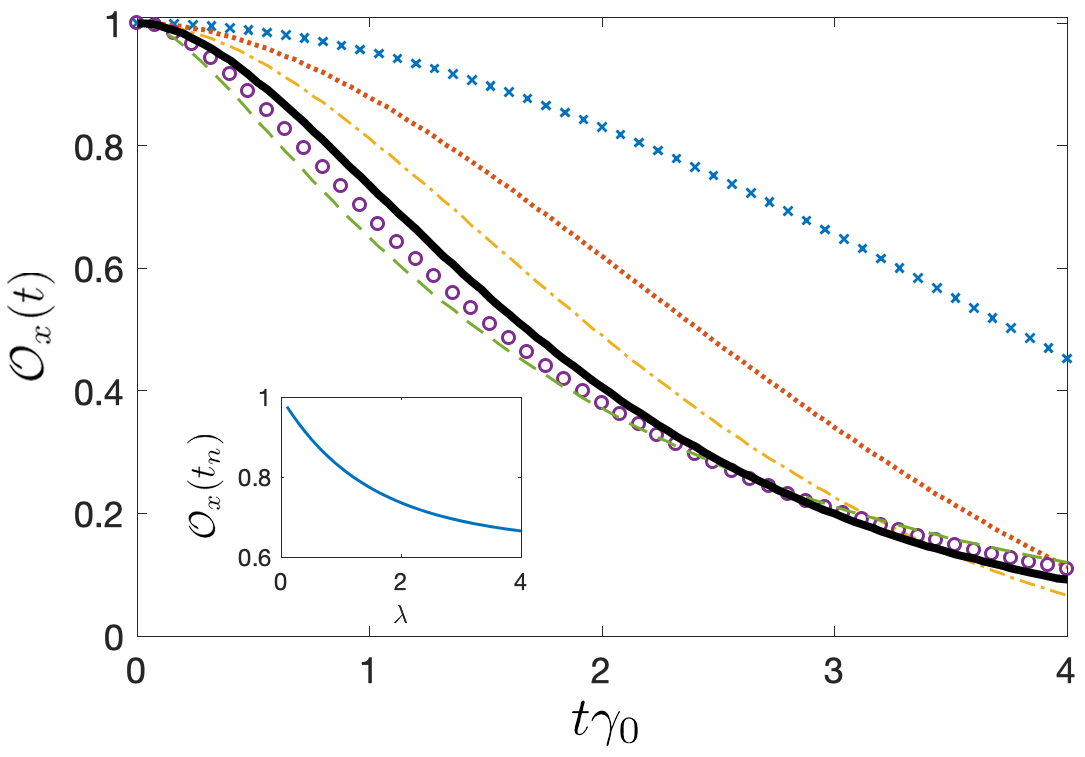}
\caption{Dynamics of the expectation value $\mathcal{O}_x(t)$ for different coupling strengths {in case of the pure AD channel, that is,} for $\lambda=0.1$ {(blue crosses)}, $\lambda=0.5$ {(red dotted line)}, $\lambda=1.0$ {(yellow dot-dashed line)}, $\lambda=3.0$ {(purple circles)}, $\lambda=5.0$ {(green dashed line)}. The thick black solid line is the separation curve between Markovian and non-Markovian dynamics, i.e., $\lambda=2.0$. In the inset, we show how {the expectation value} $\mathcal{O}_x(1/\gamma_0)$ changes with $\lambda$.}
\label{fig1}
\end{figure}

We would like to first point out that, in case of the pure channels, depending on the time $t^*$, each expectation value $\mathcal{O}_k(t)$, can have a unique correspondence with each $\lambda_n$ and $\tau_n$ for AD and PD, respectively. For illustrative purpose, we show in Fig.~\ref{fig1} the time evolution of $\mathcal{O}_x(t)$ for different values of $\lambda$ for pure AD channel. It is straightforward to note that one can find an optimal time $t_c$, (for example, in this case, around $1/\gamma_0$), at which the curves corresponding the Markovian and non-Markovian dynamics are well separated, depending on whether they are above or below the thick solid line ($\lambda=2\gamma_0$). This suggests that a single state tomography, in a well determined time $t_c$, is sufficient to estimate the degree of non-Markovianity. For example, if $t\gamma_0=1$, for each value of $\lambda$, we have a precise and distinct value of $\mathcal{O}_x(t)$. In the inset of Fig.~\ref{fig1} (assuming $t\gamma_0=1$) we show that there is an optimal region where, even for small variations in $\lambda$, the change in $\mathcal{O}_x(t_c)$ is significant. This is crucial to determine the best $t_c$ to be used in an experiment. Indeed, we need to choose a time $t_c$ that increases the accuracy of the ML algorithm but, at the same time, keep sparse the expectation values as a function of $\lambda$. For example, examining Fig.~\ref{fig1}, we see that one could choose $t\gamma_0=0.5$ but, in this case, a high precision measurement is necessary since $\mathcal{O}_x(t)$ varies not much, i.e., from approximately $0.8$ to $1.0$ as $\lambda/\gamma_0$ ranges from $0.1$ to $3.0$. This imposes a balance between the experimental precision of the measurements and the accuracy of the ML algorithm.

\begin{figure}
\centering
\includegraphics[width=0.48\textwidth]{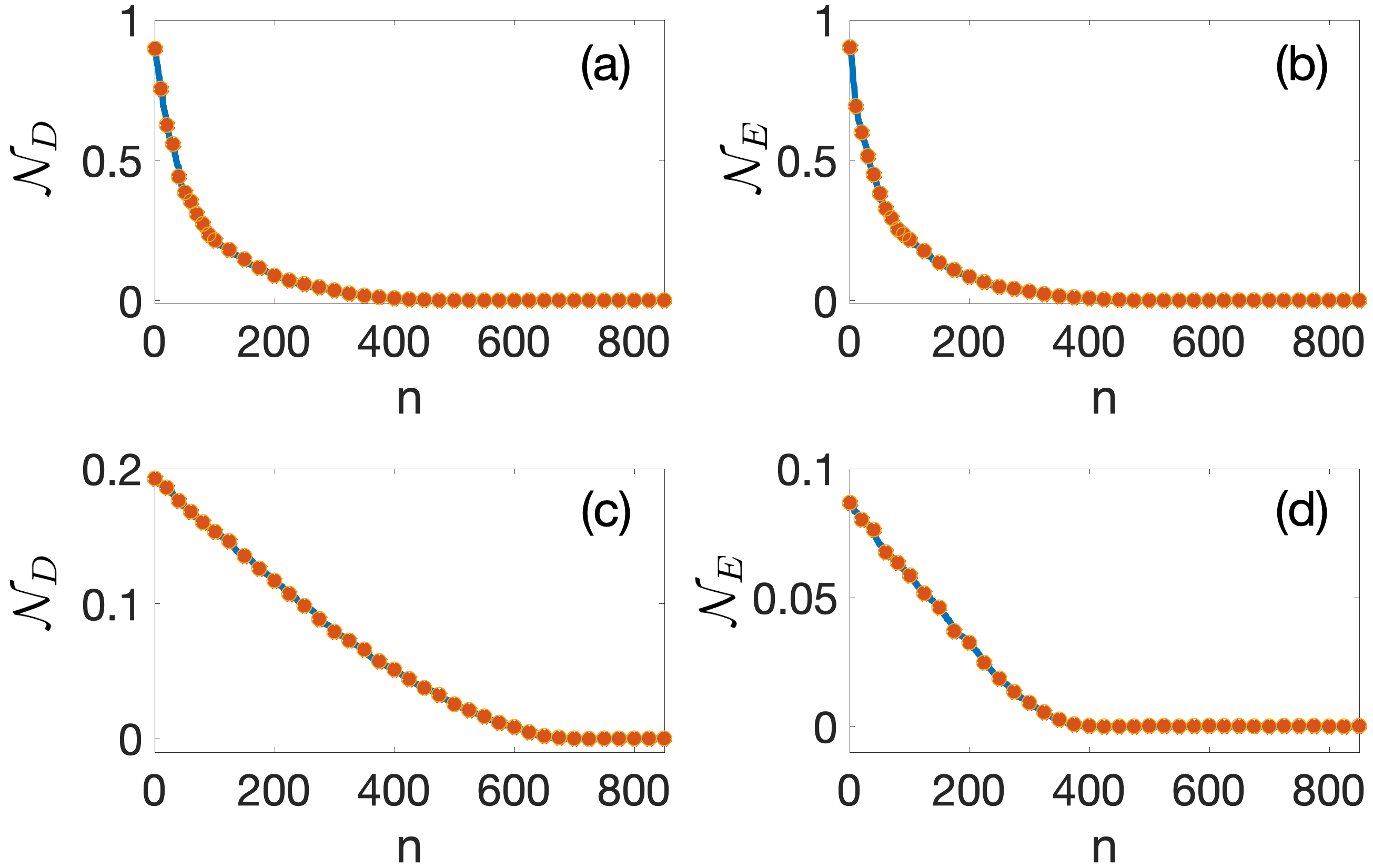}
\caption{Comparison of the estimated ({orange} circles) and theoretical values ({blue solid line}) of the degree of non-Markovianity for pure decoherence channels. The plots (a) and (b) display the results for the trace distance $\mathcal{N}_D$ and the entanglement based $\mathcal{N}_E$ measures, respectively, in case of pure AD channel. On the other hand, the plots (c) and (d) show the outcomes of the same investigation in case of pure PD channel. {The estimated values are generated by our regressor using the input data, which has not been used in training, and the target values are ordered in decreasing order for better illustration.}}
\label{fig2}
\end{figure} 
 
An important aspect of the application of ML algorithms is the concept of data normalization. Here, we also employ the procedure of feature standardization which makes the values of each feature in the dataset to have zero mean and unit variance. Such a treatment can in general speed up the algorithm convergence~\cite{ioffe2015} while increasing the accuracy of method. Thus, for each set of observables, we calculate their mean value and variance, and transform the data as \begin{equation}
\tilde{\mathcal{O}_k^n} =  (\mathcal {O}_k^n - u_k)/s_k,
\end{equation}
where $\mathcal {O}_k^n$ is a specific data of $\mathcal{O}_k$ (that is, for a particular $\lambda_n$ or $\tau_n$), $u_k$ is the mean value of the expectation $\mathcal{O}_k$ , and $s_k$ is the standard deviation of $\mathcal{O}_k$. We remark that this simple procedure can actually enhance the accuracy of the estimation up to one order of magnitude.

We now turn our attention to the results on the estimation of the degree of non-Markovianity in pure AD and PD channels, which are generated by the regressor we trained. From this point on, out of the whole database we have produced, we will keep always $70\%$ of the data (which are randomly chosen) to train the SVR, and we will reserve the remaining $30\%$ of the data to test the performance of the regressor. Note that this is a standard procedure when working with SVR, but we should also remark that the choice of these percentages can be adjusted depending on the problem to improve the prediction accuracy. In Fig.~\ref{fig2}, we show the degree of non-Markovianity predicted with our SVR model (orange circles) and the theoretical ones (blue solid line) in case of pure decoherence channels, considering both the trace distance $\mathcal{N}_D$ and entanglement $\mathcal{N}_E$ based measures of non-Markovianity. Here, in the generation of the dataset, the expectation values $\mathcal{O}_k(t)$ are calculated at the fixed time $t_c=3/\gamma_0$ ($t_c=3$) for AD (PD). We also note that the theoretical data is arranged in decreasing order and we limit the number of the estimated non-Markovianity values in the figure merely for illustrative purposes. Specifically, whereas Fig.~\ref{fig2}a and Fig.~\ref{fig2}b respectively show our findings for $\mathcal{N}_D$ and $\mathcal{N}_E$ for the AD channel, Fig.~\ref{fig2}c and Fig.~\ref{fig2}d display the results of the same analysis for the PD channel. It then becomes clear that our ML algorithm can estimate the degree of non-Markovianity with a very high precision. Indeed, the mean errors for AD and the PD channels are given by $7 \times 10^{-4}$ and $2 \times 10^{-4}$ for the trace distance measure, and $9 \times 10^{-4}$ and $9 \times 10^{-5}$ for the entanglement based measure, respectively. Therefore, for pure decoherence channels, a single tomography should be sufficient to accurately estimate the degree of memory effects.

\begin{figure}[t]
\centering
\includegraphics[width=0.48\textwidth]{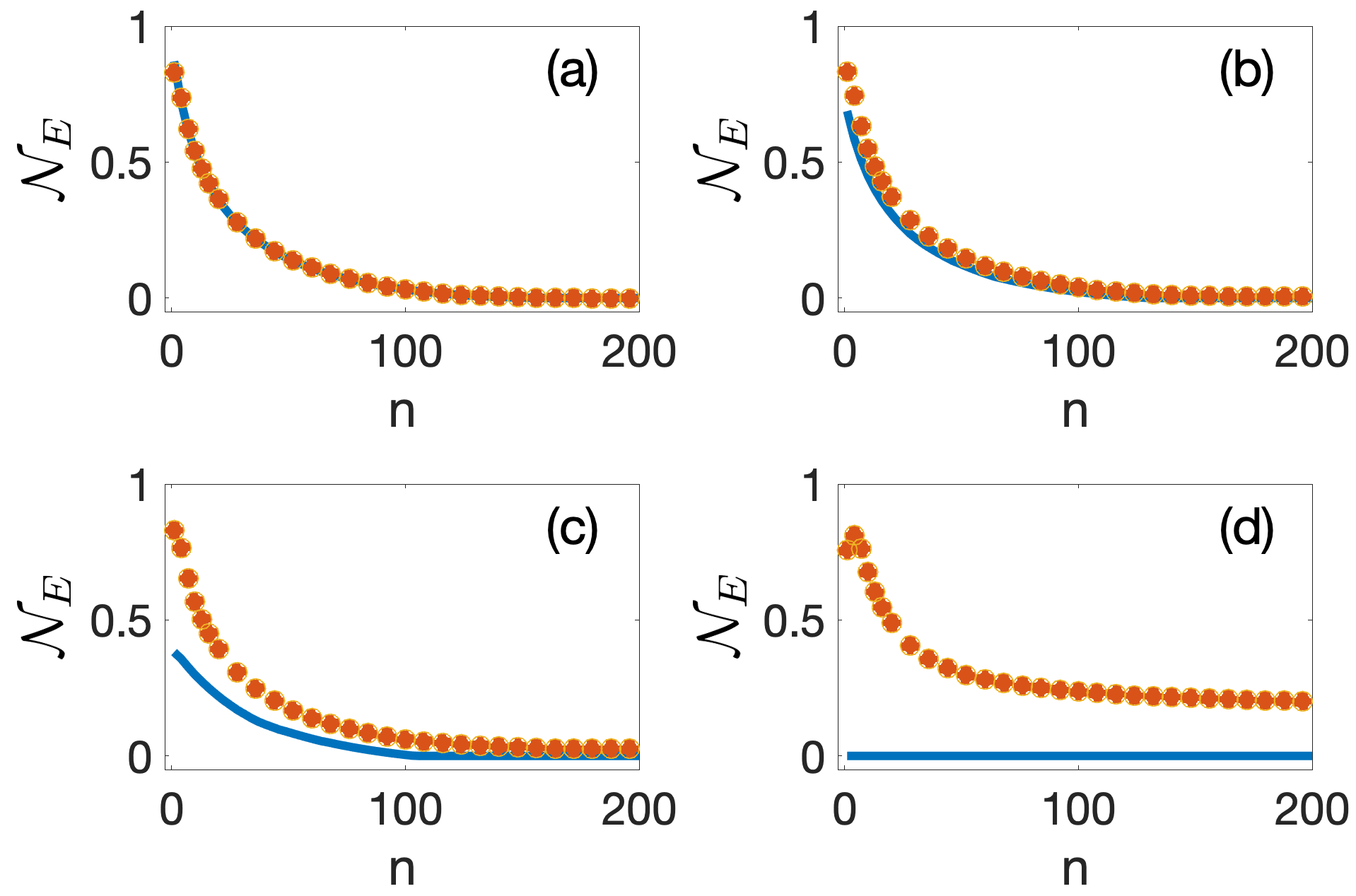}
\caption{Comparison between the predicted ({orange} circles) and theoretical values ({blue solid line}) of the degree of non-Markovianity for the AD channel considering an additional external field, as measured by the entanglement based measure $\mathcal{N}_E$, for increasing values of the field strength $\Omega$ (in units of $\gamma_0$), {that is, in (a) $\Omega=0.01$, in (b) $\Omega=0.05$, in (c) $\Omega=0.09$, and in (d) $\Omega=0.20$.} Here, our regressor has been trained with the data generated for the pure AD channel.}\label{fig3}
\end{figure}

At this point, it is important to mention that the above results on the AD channel clearly depend on the knowledge of the parameter $\gamma_0$ so that the timescale of $t_c$ can be reliably determined and our approach can be used in an experiment. If the parameter $\gamma_0$ is unknown in the considered setting, it has been recently addressed in Ref.~\cite{2005.01144} that the noise spectrum of any environment surrounding a qubit can be accurately extracted by training a deep neural network (long short-term memory network) with usual time-dynamics measurements on qubits, e.g., the two-pulse `Hahn' echo curves.

Motivated by the results we have obtained for pure decoherence channels, we would like to apply our computational approach to a natural extension of the studied problem, that is, we ask the question of what would be the consequences of an external driving field affecting the open system? This problem is certainly more involved as compared to pure decoherence since the external field induces extra oscillations in the evolution of the expectation values $\mathcal{O}$, which could be mistaken as a signature of non-Markovianity by the regressor. In this part, we choose to limit our analysis to the non-Markovianity of the AD channel quantified through the entanglement based measure $\mathcal{N}_E$. We will now assume an external driving  $\Omega\neq 0$ in Eq.~(\ref{Hamiltonian1}) and we follow the procedure that we have used to obtain the results presented in Fig.~\ref{fig2}. In fact, our first question here is: given a regressor that is trained to work with pure AD channel, how precisely is it able to estimate the degree of non-Markovianity in the presence of an external field? To answer this question we show in Fig.~\ref{fig3} the comparison between the estimated (by a regressor trained for pure AD channel) and the theoretical non-Markovianity results when the external field is non-zero for the AD channel. In the plots displayed from Fig.~\ref{fig3}a to Fig.~\ref{fig3}d, we consider the external field strength $\Omega/\gamma_0$ values to be $0.01$, $0.05$, $0.09$, and $0.20$ in respective increasing order, which in turn result in mean errors given by $1.6 \times 10^{-3}$, $2.2 \times 10^{-2}$, $6.4 \times 10^{-2}$, and $0.27$. As it can be seen comparing the predicted and theoretical non-Markovianity values, the results are satisfactory only for small perturbations, and as the driving strength increases, the regressor no longer works.

\begin{figure}[t]
\centering
\includegraphics[width=0.40\textwidth]{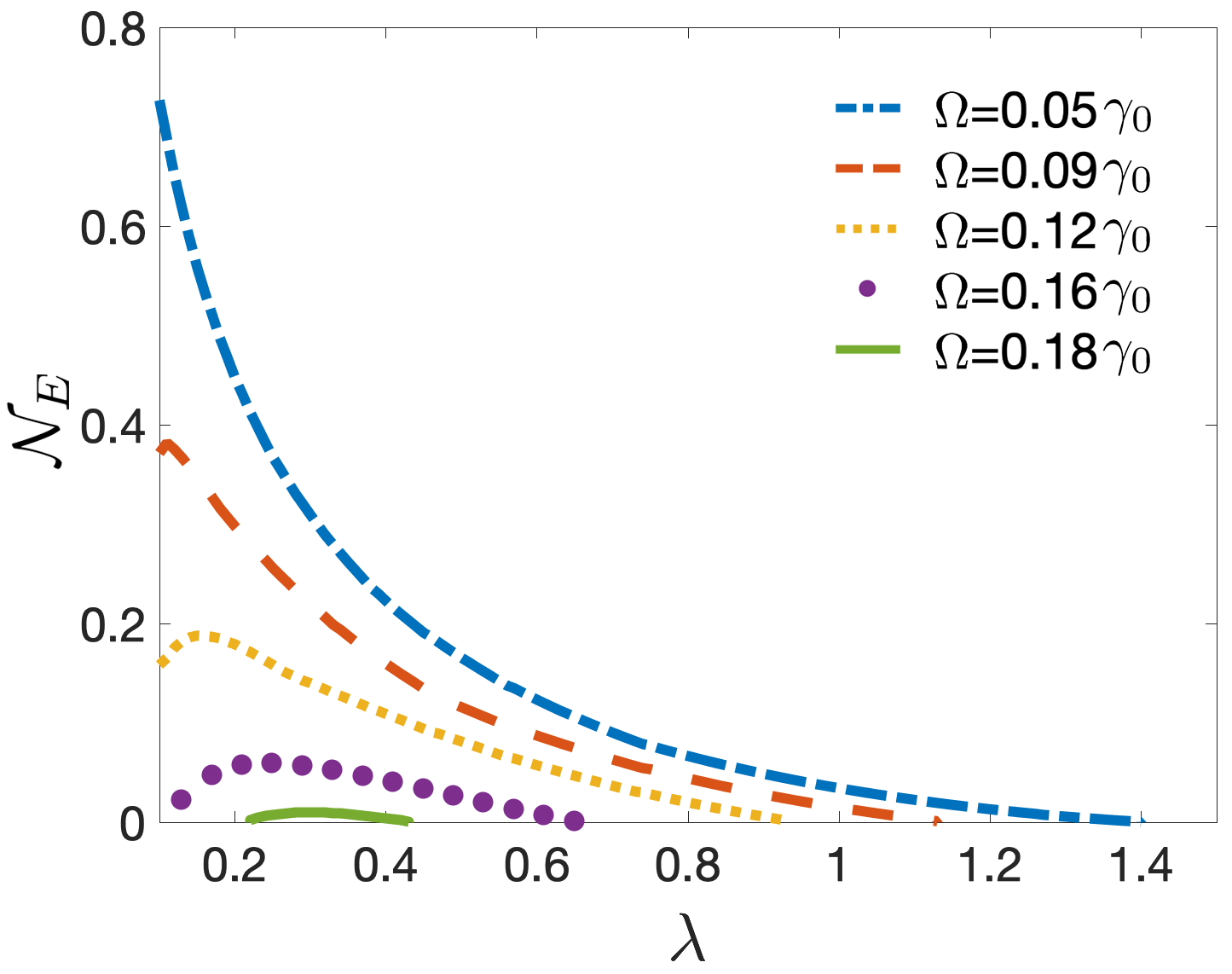}
\caption{The degree of non-Markovianity quantified by $\mathcal{N}_E$ as a function of the coupling strength $\lambda$ (in units of $\gamma_0$) for different values of external field $\Omega$ (in units of $\gamma_0$) {for the AD channel.}}
\label{fig4}
\end{figure}

Our findings in Fig.~\ref{fig3} agree with what we expected since the effects induced by the external driving can significantly alter the time evolution of the expectation values $\mathcal{O}_x(t)$, $\mathcal{O}_y(t)$, and $\mathcal{O}_z(t)$. It is also important to emphasize that revivals in the dynamics of the expectations values do not necessarily imply that the time evolution is non-Markovian. Actually, the external field $\Omega$ suppresses the memory effects in the open system dynamics despite the fact that it causes oscillations in the dynamics of the expectation values. Fig.~\ref{fig4} demonstrates this situation, i.e., while the field strength $\Omega$ increases, non-Markovianity $\mathcal{N}_E$ decreases, tending to zero even for small values of $\Omega$. This behavior is the cause of the inaccuracy of the non-Markovianity estimated by the SVR algorithm in Fig.~\ref{fig3}.

\begin{figure}[t]
\centering
\includegraphics[width=0.48\textwidth]{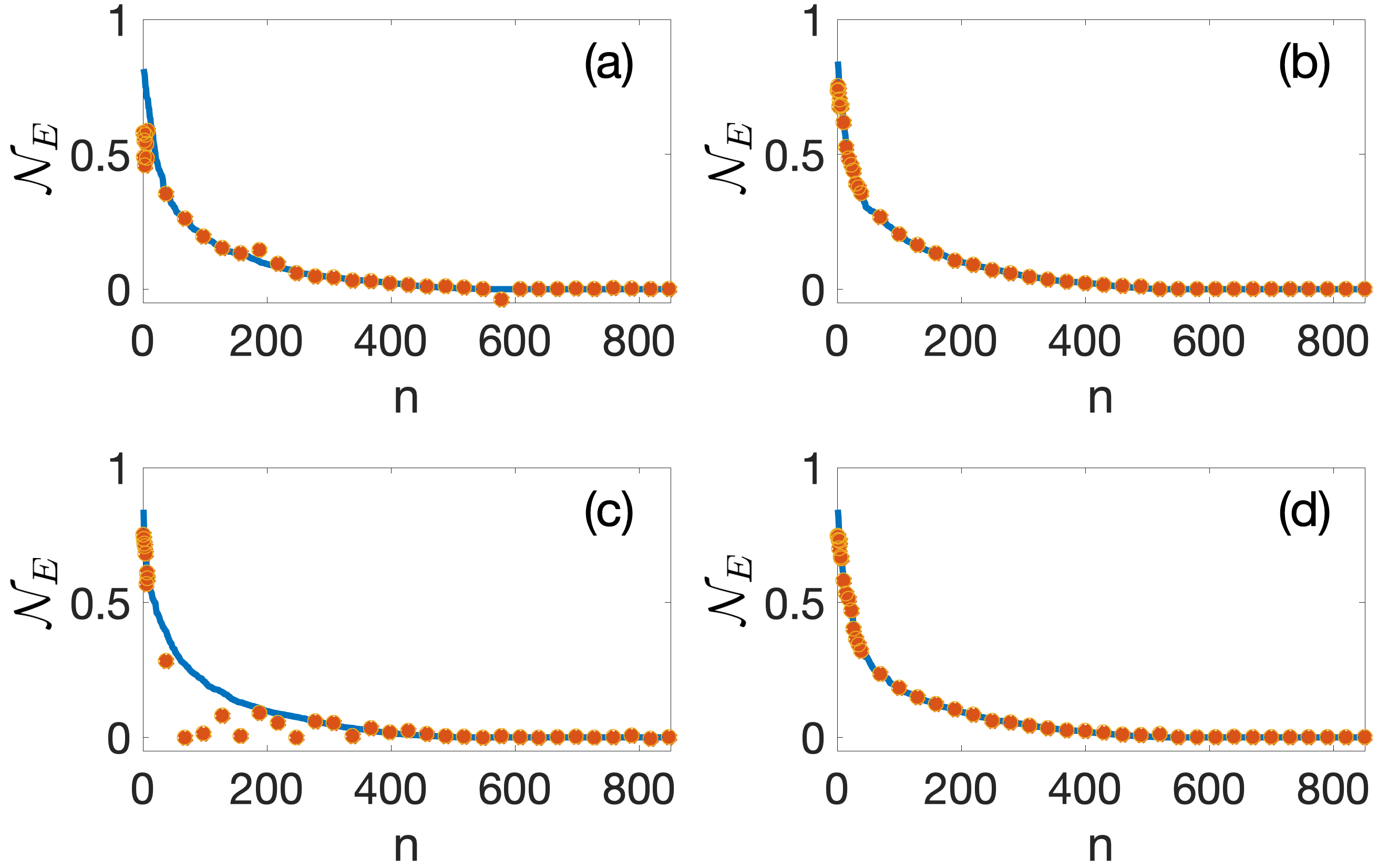}
\caption{Comparison between the estimated ({orange} circles) and theoretical values {(blue solid lines)} of the degree of non-Markovianity for the AD channel with external field, as measured by the entanglement based measure $\mathcal{N}_E$, where the regressor is trained taking into account the external field. While the plots in (a) and (c) are generated considering a single state tomography at a fixed time during the dynamics, the results in the plots (b) and (d) are obtained taking into account two tomographies {at two fixed times}.}
\label{fig5}
\end{figure}

In order to enhance the predictive power of our SVR based ML algorithm, the natural solution is to train the regressor taking into account the existence of the external field $\Omega$. Thus, we now train our algorithm assuming that the coupling strength $\lambda/\gamma_0$ takes values in the range $[0.1,3.0]$, with a step size equal to $10^{-2}$, and additionally, we consider a set of values for the drive parameter $\Omega/\gamma_0$ (ranging from $0.01$ to $0.5$), which generates a training data with $290$ samples for each $\Omega$. Here, $\Omega$ is divided with a step size equal to $10^{-2}$ for $\Omega/\gamma_0$ values between $0.01$ to $0.2$, and with a step size equal to $0.1$ between $0.2$ to $0.5$. The reason for this difference in the distribution of $\Omega$ is to have a balanced dataset, where the number of data with Markovian results is similar to that of non-Markovian ones. In Fig.~\ref{fig5}, we present the predictions of our regressor now trained in the presence of the external field. In particular, Fig.~\ref{fig5}a and Fig.~\ref{fig5}c present a comparison of the theoretical and the estimated results of the non-Markovianity measure $\mathcal{N}_E$ using the values of the expectation values $\mathcal{O}_x(t_c)$, $\mathcal{O}_y(t_c)$, and $\mathcal{O}_z(t_c)$ at fixed times $t_c=3/\gamma_0$ and $t_c=5/\gamma_0$, respectively. Note that for each case, the experimental implementation requires a single state tomography performed at time $t_c$. As can be seen from these plots, we obtain a better result for $t_c=3/\gamma_0$ as compared to $t_c=5/\gamma_0$ (mean error for these two cases are $2.6 \times 10^{-3}$ and $1.3 \times 10^{-2}$, respectively). Next, in order to further improve the estimation efficiency of our SVR algorithm, we let our regressor to have access to more information, which means that we train it using the values of $\mathcal{O}_x(t_c)$, $\mathcal{O}_y(t_c)$, and $\mathcal{O}_z(t_c)$ at two fixed times $t_{c_1}$ and $t_{c_2}$. The outcomes of our analysis in this case are shown in Fig.~\ref{fig5}b and  Fig.~\ref{fig5}d. Particularly, Fig.~\ref{fig5}b includes the results of the comparison between the estimated and the theoretical values of the non-Markovianity for the AD channel with external drive when two state tomographies are performed at times $t_{c_1}=3/\gamma_0$ and $t_{c_2}=6/\gamma_0$. On the other hand, in Fig.~\ref{fig5}d, the outcomes of the same analysis are given when the measurement times are $t_{c_1}=5/\gamma_0$ and $t_{c_2}=10/\gamma_0$. Consequently, we see that two quantum state tomographies at fixed times spaced by the intervals either $3/\gamma_0$ or $5/\gamma_0$ should be sufficient to precisely estimate the degree of non-Markovianity with mean errors $1.2 \times 10^{-3}$ and $1.3 \times 10^{-3}$, respectively.

\section{Conclusion} \label{sec6}

In summary, we have introduced an experimentally friendly approach, which utilizes ML techniques based on SVR, to estimate the degree of memory effects in the dynamics of open quantum systems. In particular, we have first considered the trace distance and entanglement based measures of non-Markovianity and demonstrated that, in case of pure AD and PD channels, a single quantum state tomography should be sufficient to estimate the value of non-Markovianity measures very precisely. Next, we have focused on AD channel but now also taking into account an external drive on the open system. We demonstrated that even though the regressor trained with pure AD data can estimate the degree of non-Markovianity relatively well for small values of the external drive strength, as the drive parameter increases, our method no longer works due to the extra oscillations induced on the expectation values $\mathcal{O}$ by the external drive. We have then shown that once our regressor is trained with the data provided by the AD channel dynamics including the external drive, it becomes once again possible to precisely estimate the degree of non-Markovianity with at most two rounds of state tomography. 

\section{Acknowledgements}
F. F. F. acknowledges support from Funda\c{c}\~{a}o de Amparo \`{a} Pesquisa do Estado de S\~{a}o Paulo (FAPESP), project number 2019/05445-7. G. K. is supported by the BAGEP Award of the Science Academy, the TUBA-GEBIP Award of the Turkish Academy of Sciences, and also by the Technological Research Council of Turkey (TUBITAK) under Grant No. 117F317. A.N. acknowledges support from Universidad Mayor through the Postdoctoral fellowship. R.C. acknowledges support from Fondecyt Iniciaci\'on No. 11180143.

\appendix
{
\section{Support Vector Machines (SVM)}\label{app}}

In this part, we intend to first present an intuitive explanation of the SVM based algorithms in a simple setting, and then we will provide the mathematical details of this approach. We remark that SVM can be used for regression (SVR) and for classification (SVC). Indeed, we perform regression analysis using the SVR algorithm throughout our study rather than working on a classification task via SVC. That is to say that, we actually aim to determine the degree of non-Markovianity of a given open quantum system dynamics rather than only classifying them as being Markovian or non-Markovian. However, due to the fact that SVM algorithm has been originally introduced for classification problems, we start our elaboration of support vector based algorithms considering a simple classification problem. The reason for this is two fold. First, approaches to classification and regression tasks using support vectors share a lot of common properties despite some differences in their applications. Second, we think that it is more illustrative to discuss classification before regression for pedagogical purposes.

{
\subsection{An instructive SVC Example}}

In order to elucidate the main idea involved in the classification procedure using SVC, we would like start by discussing a simple example. Assume that we focus on the AD model with an external field $\Omega$ studied in the main text, and consider the model parameter $\lambda$ as the \textit{feature} and the degree of non-Markovianity $\mathcal{N}$ as the \textit{target} value. We should first emphasize that the parameter $\lambda$ is not experimentally accessible in general. In fact, in our analysis in the main text, we have used the expectation values of spin observables that can be measured in experiments as the features in our algorithm. However, for the simplicity of illustration, here we suppose that $\lambda$ is our single input value for the SVC algorithm.

\begin{figure}
\centering
\includegraphics[width=0.38\textwidth]{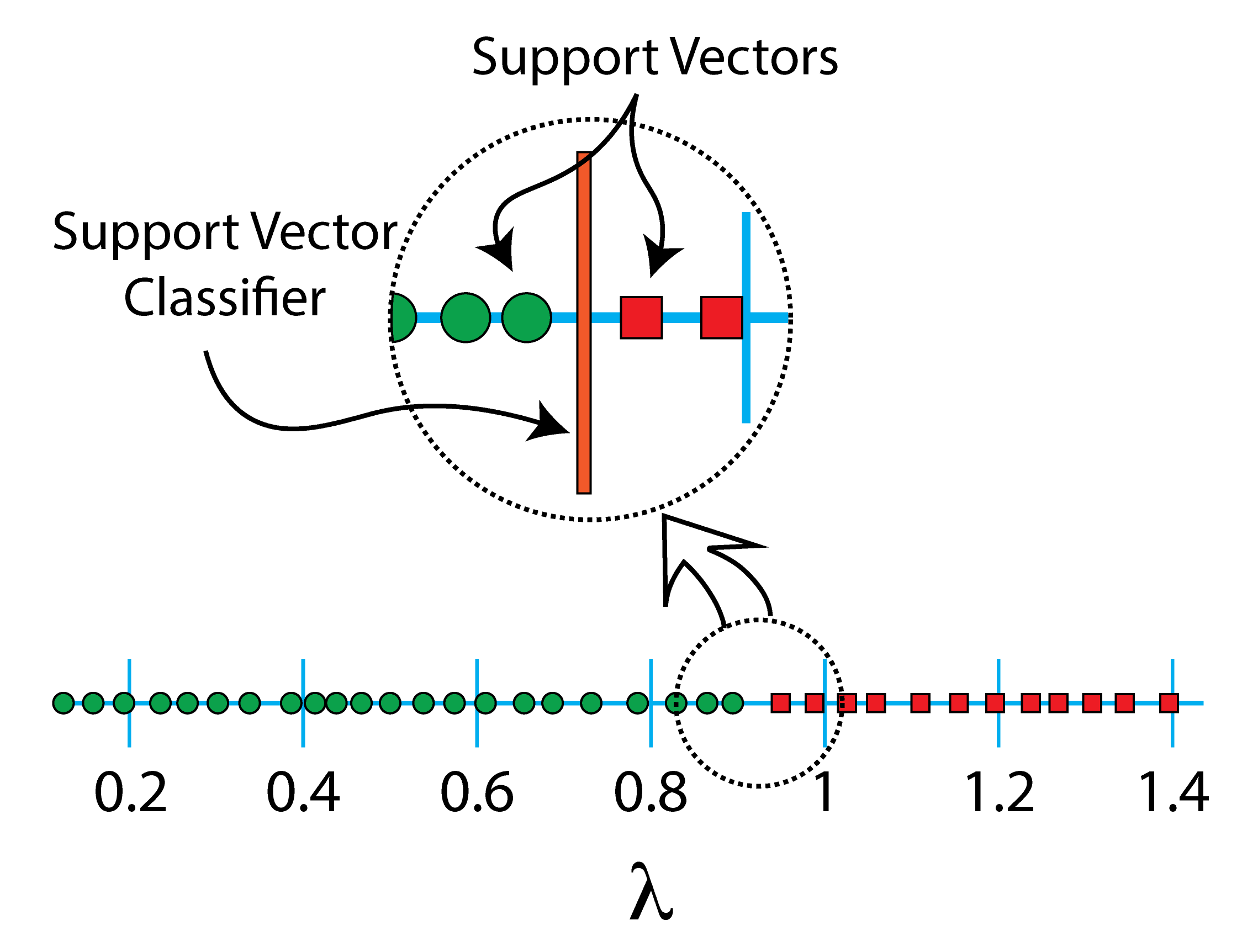}
\caption{Simple illustration of a support vector classifier and along with its support vectors in a one dimensional problem. The parameter $\lambda$ (in units of $\gamma_0$) in the AD dynamics is supposed to be used as a feature where $\Omega=0.12\gamma_0$, and while the green circles denote non-Markovian observations, the red squares show Markovian ones.}
\label{fig6}
\end{figure}

Let us now look at the case $\Omega=0.12\gamma_0$ in Fig.~\ref{fig4} where the degree of non-Markovianity $\mathcal{N}_E$ is displayed as a function of the model parameter $\lambda$. In this case, open system dynamics is Markovian only for $\lambda\gtrapprox0.95 \gamma_0$. Therefore, it should be possible to determine a simple boundary separating two classes as Markovian and non-Markovian, which makes the problem rather trivial. Indeed, choosing our input as a single parameter, namely $\lambda$ here, we limit ourselves to a one dimensional classification problem. In Fig.~\ref{fig6}, we display a simple illustration of the problem where we highlight non-Markovian evolutions with green circles (before $\lambda \approx 0.95\gamma_0$) and Markovian processes with red squares (after $\lambda \approx 0.95\gamma_0$). The main idea behind the SVC algorithm is to define a threshold that is able to divide the two classes in a reliable way. The shortest distance between the observations at the boundary of the class and the classifier threshold is called the \textit{margin} and the purpose of the algorithm is to simply maximize it. As can be seen in Fig.~\ref{fig6}, it is the support vector classifier that will define the division between the classes, while the support vectors define the margins that are to be maximized. It is worth to note that if the data is presented in a one (two) dimensional system then the classifier is a point (line). In fact, the classifier will be in general an hyperplane which has one dimension less than the number of dimensions that define the class, which is determined by the number of features. This is also the reason why we actually supposed to use the model parameter $\lambda$ as the single input of the algorithm in this part to be able to provide an instructive graphical illustration.

\begin{figure}[t]
\centering
\includegraphics[width=0.39\textwidth]{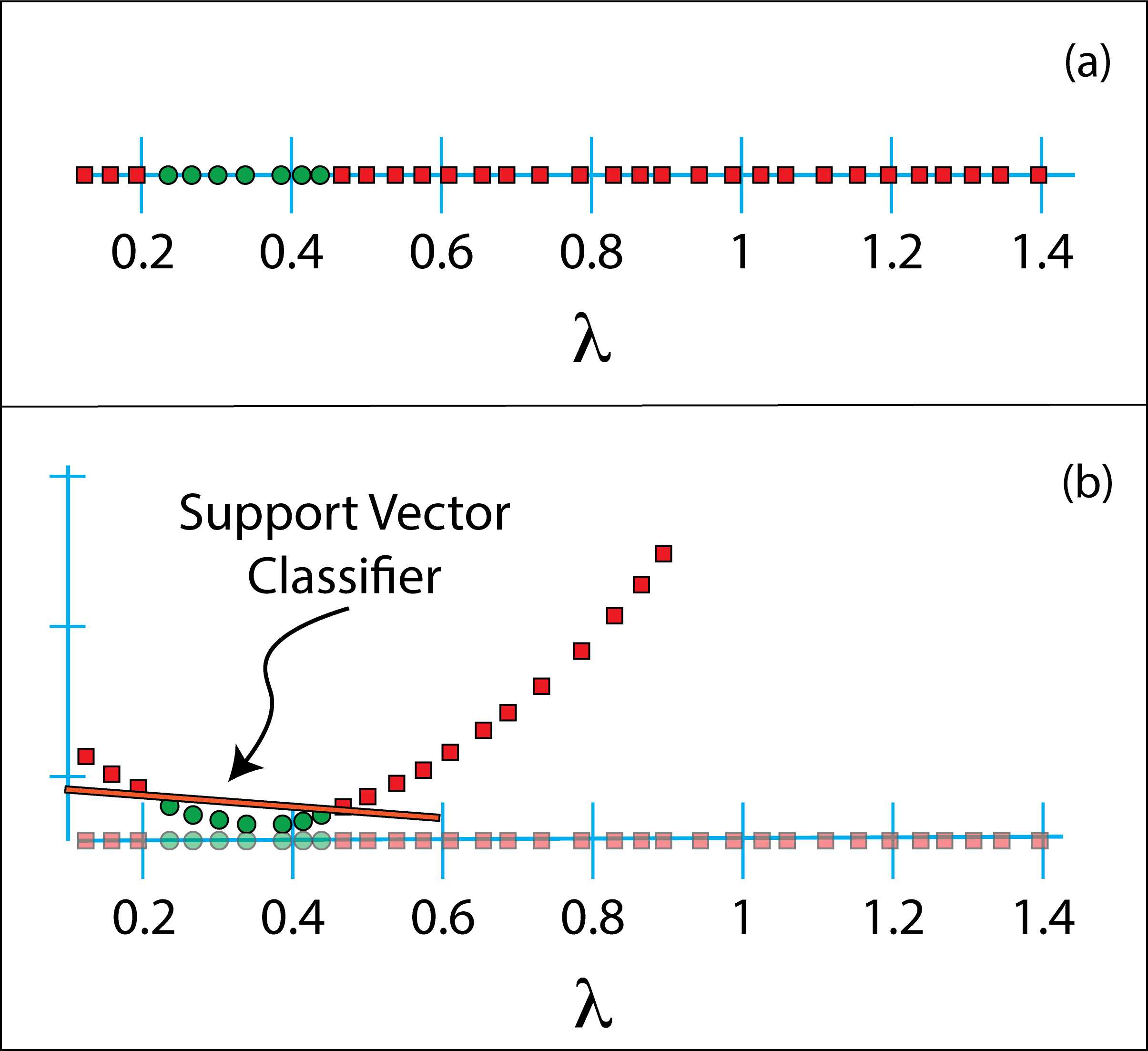}
\caption{Illustration of a kernel transformation. In (a) we assume that the parameter $\lambda$ in the AD dynamics is supposed to be used as a feature where $\Omega=0.18\gamma_0$, and we show the data in one dimension before the kernel transformation, in which case linear separation is impossible. In (b) we display the data on a two dimensional space after the polynomial kernel transformation of order 2, which makes possible to linearly separate the data with a line, allowing for an efficient application of the SVC algorithm. As the green circles denote non-Markovian observations, the red squares show Markovian ones.}
\label{fig7}
\end{figure}

Given the above explanation in a rather trivial case, where the linear separation of the classes is possible, a natural question arises about how one can use SVC when the linear separation of the data is impossible. For instance, let us have a look at the case where $\Omega=0.18\gamma_0$ in Fig.~\ref{fig4}. It is now clearly not possible to find a point that can efficiently divide the dynamics as being non-Markovian and Markovian, as demonstrated in Fig.~\ref{fig7}a, since the time evolution becomes non-Markovian only when $0.2 \lessapprox\lambda/\gamma_0 \lessapprox 0.42$. SV algorithms are still very effective in such cases but before the usual machinery of the algorithm one needs to apply a data transformation also known as a kernel transformation. In our calculations mentioned in the main text, we have used the well-known radial basis function (RBF) kernel but, here in this part, we consider a polynomial kernel of order 2 to better explain the main idea since it allows for the graphical visualization of the process. The polynomial kernel transformation can be written as $k(m,n) = (m\cdot n+r)^2$ where $m$ and $n$ refer to any two observations in the sample and $r$ is the coefficient of the polynomial. In this case, it is straightforward to see that $(m\cdot n+r)^2=\vec{u}\cdot\vec{v}$, where $\vec{u}=(\sqrt{2r}m,m^2,r)$ and $\vec{v} = (\sqrt{2r}n,n^2,r)$. This implies that each sample ($m$ and $n$) is now expanded into a three-dimensional or even two-dimensional vector if we ignore the component $z$ which is identical for both vectors. Thus, each observation is now described in a two dimensional space as shown in Fig.~\ref{fig7}b rather than a one dimensional space as displayed in Fig.~\ref{fig7}a. To put it simply, the kernel transformation extends the data in a higher dimensional space so that the linear separation is possible and thus the SVM methodology can be utilized to accurately. In case of the RBF kernel, that we have used in the main text for regression, unlike the above polynomial kernel, the expansion is multidimensional and the support vector classifier is a multidimensional hyperplane. Explicitly, the RBF kernel transformation is given by
\begin{equation}
k(m,n) =\exp(-||m-n||^2)/2\sigma^2,
\end{equation}
where $||\cdot ||^2$ is the squared Euclidean distance and $\sigma>0$ defines the tolerance in the limits of decision. Also note that the multidimensional character of the RBF kernel transformation comes from the infinite series expansion of the exponential. 

In the first section of the appendix, we aimed to present a brief intuitive explanation of the SV based classification algorithm. Indeed, SV based algorithms can also be used for regression tasks as is demonstrated with our analysis in the main text. In the following, we intend to provide mathematical details of these algorithms, which share a lot common concepts in their applications, such as the determination of the support vectors, margins and kernel transformations.

\begin{figure*}[t]
\centering
\includegraphics[width=0.79\textwidth]{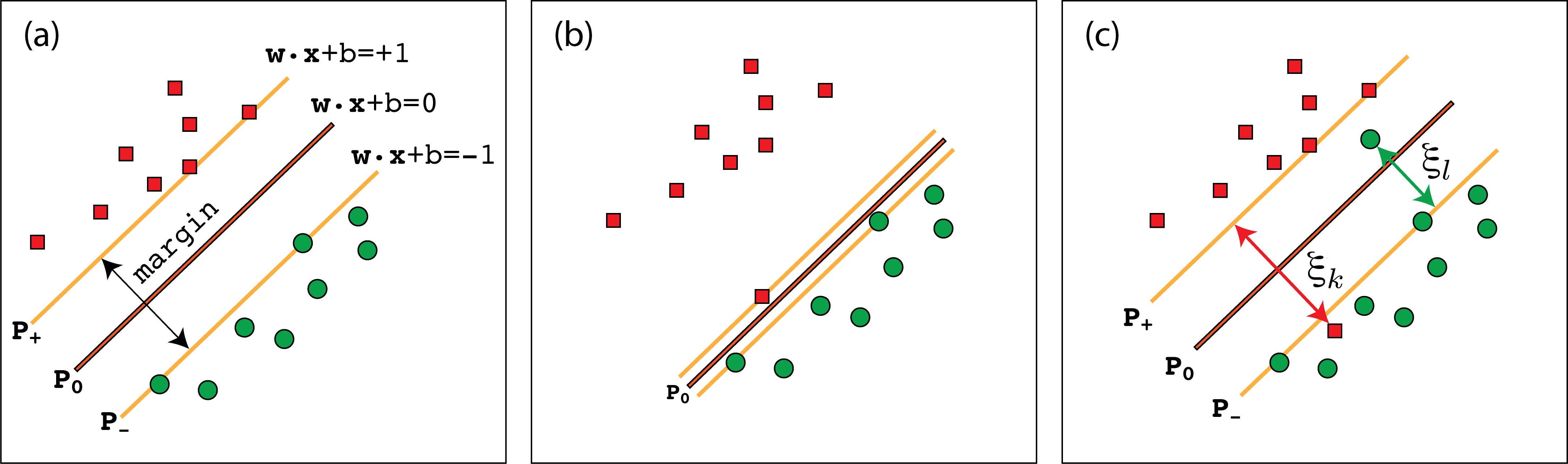}
\caption{In (a) we present a schematic representation of a classifier (hyperplane) separating the data, where the task of the algorithm is to maximize the margin defined by the support vectors. In (b) we show an example considering a training data where a $\textit{hard margin}$ fails to efficiently separate the classes. In (c) we demonstrate how the case in (b) can be improved using a \textit{soft margin} by allowing some error.}
\label{fig8}
\end{figure*}

{
\subsection{Mathematical Details of SV based Algorithms}}

\subsubsection{Classification}

As mentioned previously, the first step of the algorithm is the data transformation so that it becomes possible to find an optimal hyperplane linearly separating the data into two classes. In our problem, we make use of the expectation values of three spin observables at different times as features $\textbf{x}_i$ to estimate the non-Markovian character of dynamics. In case of classification, the target parameter can naturally assume only two values, that is, $y_i=+1$ (Markovian) or $y_i=-1$ (non-Markovian). Let us introduce two hyperplanes defined by 
\begin{eqnarray}
\textbf{w}\cdot\textbf{x}_i +b &=& + 1\;\;\;\;{\rm when}\;\;\;\;y_i=+1,\nonumber\\
\textbf{w}\cdot\textbf{x}_i +b &=& - 1\;\;\;\;{\rm when}\;\;\;\;y_i=-1,\label{cond}
\end{eqnarray}
as graphically demonstrated in Fig.~\ref{fig8}a, where $\textbf{w}$ and $b$ are the fitting parameters. In particular, these two planes should separate the whole space and define a margin, i.e., an empty region described by $-1 \leq (\textbf{w}\cdot\textbf{x}_i +b) \leq + 1$. The main task here is to determine a linear separator with the largest possible margin, at the same time, making sure that there are no data points left between the $P_{-}=\textbf{w}\cdot\textbf{x}_i +b = - 1$ and $P_+=\textbf{w}\cdot\textbf{x}_i +b = + 1$ hyperplanes shown in Fig.~\ref{fig8}a. In other words, since the distance between these two planes is proportional to $1/\left\Vert \textbf{w}\right\Vert$, the problem actually boils down to numerically determine $\textbf{w}$ and $b$ which minimize $\left\Vert \textbf{w}\right\Vert$ (or equivalently $\frac{1}{2} \left\Vert \textbf{w}\right\Vert ^2$) while making sure that the above mentioned constraints are satisfied.
 
Up until this point, we have considered a somewhat ideal scenario where the described procedure can be implemented in a quite straightforward manner. However, for certain training data sets, it is not the most ideal approach. For example, suppose that after the kernel transformation, we end up with a training data set as shown in Fig.~\ref{fig8}b. In such a case, although the hyperplane $P_0$ separates the two classes perfectly, it clearly is not the most efficient one. Indeed, the absence of flexibility in the algorithm to allow for some errors, might result in the generation of inefficient hyperplanes for the dividing the data set. In order to avoid this issue, Cortes and Vapnik proposed the concept of a soft margin which would allow for a certain degree of error in classification to keep the margin as wide as possible~\cite{svm95}. According to their approach, the constraints are modified as follows
\begin{eqnarray}
\textbf{w}\cdot\textbf{x}_i +b &\geq& + 1-\xi_i\;\;\;\;{\rm when}\;\;\;\;y_i=+1,\nonumber\\
\textbf{w}\cdot\textbf{x}_i +b &\leq& - 1+\xi_i\;\;\;\;{\rm when}\;\;\;\;y_i=-1,\label{cond2}
\end{eqnarray}
where $\xi_i$ are slack variables which measures the distance of $\textbf{x}_i$ from the corresponding margin if $\textbf{x}_i$ is on the wrong side of the margin, otherwise they are zero (see Fig.~\ref{fig8}c).

As a certain amount of error is now allowed in the algorithm, it is necessary to introduce a slightly different criterion to determine the hyperplanes. Therefore, rather than maximizing the distance between the hyperplanes $P_{-}$ and $P_{+}$ as in the previous case (or in other words, minimizing $\frac{1}{2} \left\Vert \textbf{w}\right\Vert ^2$), one now requires to minimize the following quantity:
\begin{equation}
\frac{1}{2} \left\Vert \textbf{w}\right\Vert ^2 + C\sum_i \xi_i,\label{cond3}
\end{equation}
under the constraints given in Eq.~(\ref{cond2}), where, as the margin is being maximized, an extra term $C$ is introduced to take into account a penalty for possible errors. Here, the new positive parameter $C$ actually weighs the degree of allowed errors in the algorithm. More explicitly, $C$ is a trade-off parameter whose value determines whether one wants a better classification of training data impairing the wide margin.

\subsubsection{Regression}

As mentioned earlier, SVM algorithm has been originally developed for the purpose of classification. Later on, it has been extended to work as a regressor~\cite{Drucker96,Drucker}. As in the case of linear regression algorithms, SV based regression algorithm intends to find a line (or in general, an hyperplane) that fits training points while minimizing possible errors. In other words, the task of the SVR algorithm is to find a continuous linear function $f(\textbf{x})= \textbf{w}\cdot\textbf{x} + b$  which approximates the mapping from an input data (features) to real numbers (target) in accordance with the training data set. A kernel transformation should still be performed to linearize the training data in a higher dimensional parameter space. In fact, the approach here is actually similar to the one presented for the classification problem except for a few important differences. Most importantly, in case of the SV classifier, since the aim is to efficiently separate the classes, a decision boundary is formulated keeping the margin as wide as possible between the data points closest to it (support vectors), where some data points are allowed to fall inside this margin (See Fig.~\ref{fig8}).

\begin{figure}[t]
\centering
\includegraphics[width=0.32\textwidth]{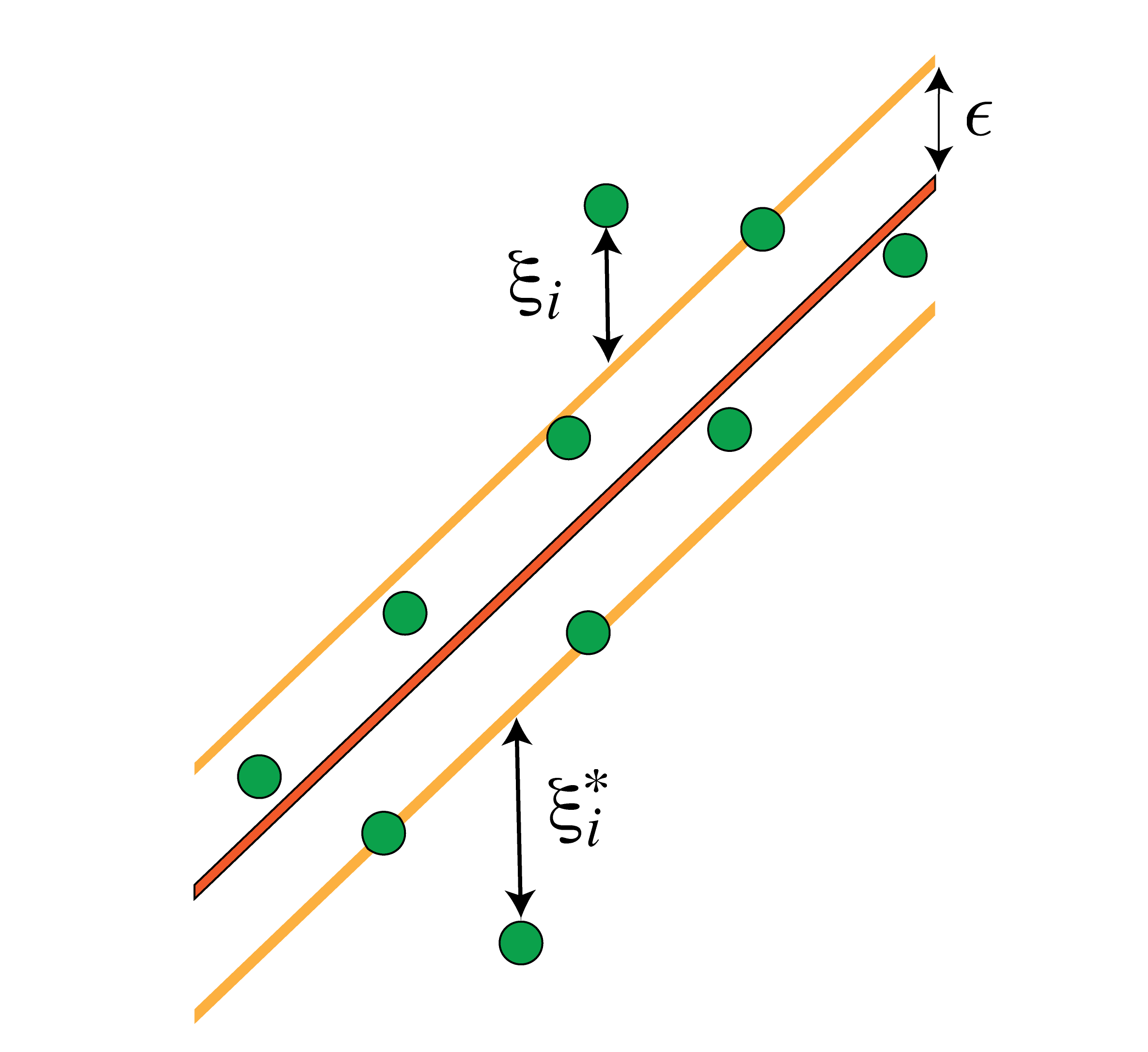}
\caption{Schematic representation of a SV based algorithm which is designed to work in a regression task. The main idea is to utilize the hyperplane shown by red solid line in the middle as a regression function, at the same time, introducing a certain degree of penalty (through slack variables $\xi_i$ and $\xi_i^\ast$) for the training data set falling outside the margin, which is defined by the parameter $\epsilon$.}
\label{fig9}
\end{figure}

When it comes to the SV based regression algorithm, a margin still needs to be fixed but with a completely different purpose. In order to define this margin, we need to introduce an additional parameter $\epsilon$ whose value is to be decided based on the distribution of the training data. This new parameter defines a region around the function $f(\textbf{x})$ that will be optimized to keep the deviation from the training data set minimal. Thus, the minimization constraints now become
\begin{eqnarray}
 -\epsilon \leq y_i - \textbf{w}\cdot\textbf{x}_i -b \leq \epsilon \label{cond4}.
\end{eqnarray}
However, it is quite possible that no such linear function $f(\textbf{x})$ exists that satisfy the above constraints. For this reason, non-negative slack variables $\xi_i$ and $\xi_i^\ast$ are introduced to the model similarly to the soft margin concept in SV based classification. As graphically displayed in Fig.~\ref{fig9}, the parameter $\epsilon$ describes a region of space around the hyperplane, inside which the deviations are ignored (the so-called $\epsilon$-insensitive margin) and the instances outside this limit are penalized with the help of slack variables $\xi_i$ and $\xi_i^\ast$, allowing for a certain degree of regression errors. With the above mentioned considerations, the problem that needs to be dealt with for efficiently training the regressor is now the minimization of the following quantity
\begin{eqnarray}
\frac{1}{2} \left\Vert \textbf{w}\right\Vert ^2 + C\sum_i \left( \xi_i + \xi_i^\ast \right),
\end{eqnarray}
under the constrains given by
\begin{eqnarray}
 -(\epsilon + \xi_i^\ast ) \leq y_i - \textbf{w}\cdot\textbf{x}_i -b \leq \epsilon + \xi_i
\end{eqnarray}
where the positive valued parameter $C$ controls the penalty imposed on the training data lying outside the $\epsilon$-insensitive margin and helps prevent overfitting of the data. The above problem is clearly equivalent to the constrained optimization task described by Eq.~(\ref{min01}) and Eq.~(\ref{min02}) in the main text. At this point, it is also important to note that the process of maximizing margins, involved in the process of minimizing the quantity $\left\Vert \textbf{w}\right\Vert ^2$, has a different role in the case of the regression algorithm as compared to the classifier, since the width of the margin is limited by the parameter $\epsilon$ in the former approach. We should also emphasize that the choice of the parameter $\epsilon$ is crucial for the efficiency of the algorithm as it defines the maximum possible error allowed for the regressor.

{
\subsection{Numerical Implementation of SVR}}

Having provided above a rather technical description of SV based regression approach that we have used in our analysis, we would like to finally give some details about the numerical implementation of the training and regression procedures. In our work, the SVR algorithm has been implemented using the scikit-learn platform~\cite{scikit-learn} which is a well-recognized free machine learning library for the Pyhton programming language. In fact, SVR implementation in scikit-learn itself is internally handled using the libsvm library~\cite{libsvm}. Within this framework, the optimization problem described by Eq.~(\ref{min01}) and Eq.~(\ref{min02}) in the main text is numerically solved with a sequential minimal optimization type decomposition technique proposed by Fan et al. in Ref.~\cite{Fan2005}. The free parameters in the considered SVR algorithm are $\epsilon$ and $C$ that respectively represent the size of the epsilon-tube within which errors are ignored, and the tolerance for deviations larger than $\epsilon$. In addition, different kernels can also be considered depending on the distribution of the training data. In our study, we have carried a careful analysis performing numerous implementations of the algorithm to achieve the best accuracy. Accordingly, we have used the previously defined RBF kernel and set the model parameters as $\epsilon=10^{-3}$, $C=1.0$, and the tolerance for the stopping criterion as $10^{-3}$.

Lastly, as we use the expectation values of the three spin operators $\mathcal{O}_x$, $\mathcal{O}_y$ and $\mathcal{O}_z$ at one or at most two time points during the open system dynamics as features, and the degree of non-Markovianity $\mathcal{N}$ as the target value, our training data in general is structured as a set of arrays whose content read
\begin{equation}
[\mathcal{N}, \mathcal{O}_x(t_1), \mathcal{O}_y(t_1), \mathcal{O}_z(t_1), \mathcal{O}_x(t_2), \mathcal{O}_y(t_2), \mathcal{O}_z(t_2)],
\end{equation}
having one target and six feature values. Here, each array has a different content since both expectations values at different times and the degree of non-Markovianity are calculated considering different model parameters, such as the external field strength $\Omega$ and the coupling constant $\lambda$. We also stress that we have used $70\%$ of the data for training the regression algorithm. The remaining $30\%$ of the whole data set, which was not used in training, has been kept for checking the accuracy of the model to predict the desired target value.

\bibliography{references}

\begin{thebibliography}{78}%
\makeatletter
\providecommand \@ifxundefined [1]{%
 \@ifx{#1\undefined}
}%
\providecommand \@ifnum [1]{%
 \ifnum #1\expandafter \@firstoftwo
 \else \expandafter \@secondoftwo
 \fi
}%
\providecommand \@ifx [1]{%
 \ifx #1\expandafter \@firstoftwo
 \else \expandafter \@secondoftwo
 \fi
}%
\providecommand \natexlab [1]{#1}%
\providecommand \enquote  [1]{``#1''}%
\providecommand \bibnamefont  [1]{#1}%
\providecommand \bibfnamefont [1]{#1}%
\providecommand \citenamefont [1]{#1}%
\providecommand \href@noop [0]{\@secondoftwo}%
\providecommand \href [0]{\begingroup \@sanitize@url \@href}%
\providecommand \@href[1]{\@@startlink{#1}\@@href}%
\providecommand \@@href[1]{\endgroup#1\@@endlink}%
\providecommand \@sanitize@url [0]{\catcode `\\12\catcode `\$12\catcode
  `\&12\catcode `\#12\catcode `\^12\catcode `\_12\catcode `\%12\relax}%
\providecommand \@@startlink[1]{}%
\providecommand \@@endlink[0]{}%
\providecommand \url  [0]{\begingroup\@sanitize@url \@url }%
\providecommand \@url [1]{\endgroup\@href {#1}{\urlprefix }}%
\providecommand \urlprefix  [0]{URL }%
\providecommand \Eprint [0]{\href }%
\providecommand \doibase [0]{http://dx.doi.org/}%
\providecommand \selectlanguage [0]{\@gobble}%
\providecommand \bibinfo  [0]{\@secondoftwo}%
\providecommand \bibfield  [0]{\@secondoftwo}%
\providecommand \translation [1]{[#1]}%
\providecommand \BibitemOpen [0]{}%
\providecommand \bibitemStop [0]{}%
\providecommand \bibitemNoStop [0]{.\EOS\space}%
\providecommand \EOS [0]{\spacefactor3000\relax}%
\providecommand \BibitemShut  [1]{\csname bibitem#1\endcsname}%
\let\auto@bib@innerbib\@empty
%</preamble>
\bibitem [{\citenamefont {Jordan}\ and\ \citenamefont
  {Mitchell}(2015)}]{jordan255}%
  \BibitemOpen
  \bibfield  {author} {\bibinfo {author} {\bibfnamefont {M.~I.}\ \bibnamefont
  {Jordan}}\ and\ \bibinfo {author} {\bibfnamefont {T.~M.}\ \bibnamefont
  {Mitchell}},\ }\bibfield  {title} {\enquote {\bibinfo {title} {Machine
  learning: Trends, perspectives, and prospects},}\ }\href {\doibase
  10.1126/science.aaa8415} {\bibfield  {journal} {\bibinfo  {journal}
  {Science}\ }\textbf {\bibinfo {volume} {349}},\ \bibinfo {pages} {255--260}
  (\bibinfo {year} {2015})}\BibitemShut {NoStop}%
\bibitem [{\citenamefont {Pedregosa}\ \emph {et~al.}(2011)\citenamefont
  {Pedregosa}, \citenamefont {Varoquaux}, \citenamefont {Gramfort},
  \citenamefont {Michel}, \citenamefont {Thirion}, \citenamefont {Grisel},
  \citenamefont {Blondel}, \citenamefont {Prettenhofer}, \citenamefont {Weiss},
  \citenamefont {Dubourg}, \citenamefont {Vanderplas}, \citenamefont {Passos},
  \citenamefont {Cournapeau}, \citenamefont {Brucher}, \citenamefont {Perrot},\
  and\ \citenamefont {{{\'E}}douard Duchesnay}}]{scikit-learn}%
  \BibitemOpen
  \bibfield  {author} {\bibinfo {author} {\bibfnamefont {Fabian}\ \bibnamefont
  {Pedregosa}}, \bibinfo {author} {\bibfnamefont {Ga{{\"e}}l}\ \bibnamefont
  {Varoquaux}}, \bibinfo {author} {\bibfnamefont {Alexandre}\ \bibnamefont
  {Gramfort}}, \bibinfo {author} {\bibfnamefont {Vincent}\ \bibnamefont
  {Michel}}, \bibinfo {author} {\bibfnamefont {Bertrand}\ \bibnamefont
  {Thirion}}, \bibinfo {author} {\bibfnamefont {Olivier}\ \bibnamefont
  {Grisel}}, \bibinfo {author} {\bibfnamefont {Mathieu}\ \bibnamefont
  {Blondel}}, \bibinfo {author} {\bibfnamefont {Peter}\ \bibnamefont
  {Prettenhofer}}, \bibinfo {author} {\bibfnamefont {Ron}\ \bibnamefont
  {Weiss}}, \bibinfo {author} {\bibfnamefont {Vincent}\ \bibnamefont
  {Dubourg}}, \bibinfo {author} {\bibfnamefont {Jake}\ \bibnamefont
  {Vanderplas}}, \bibinfo {author} {\bibfnamefont {Alexandre}\ \bibnamefont
  {Passos}}, \bibinfo {author} {\bibfnamefont {David}\ \bibnamefont
  {Cournapeau}}, \bibinfo {author} {\bibfnamefont {Matthieu}\ \bibnamefont
  {Brucher}}, \bibinfo {author} {\bibfnamefont {Matthieu}\ \bibnamefont
  {Perrot}}, \ and\ \bibinfo {author} {\bibnamefont {{{\'E}}douard
  Duchesnay}},\ }\bibfield  {title} {\enquote {\bibinfo {title} {Scikit-learn:
  Machine learning in python},}\ }\href
  {http://jmlr.org/papers/v12/pedregosa11a.html} {\bibfield  {journal}
  {\bibinfo  {journal} {Journal of Machine Learning Research}\ }\textbf
  {\bibinfo {volume} {12}},\ \bibinfo {pages} {2825--2830} (\bibinfo {year}
  {2011})}\BibitemShut {NoStop}%
\bibitem [{\citenamefont {Dunjko}\ and\ \citenamefont
  {Briegel}(2018)}]{Dunjko_2018}%
  \BibitemOpen
  \bibfield  {author} {\bibinfo {author} {\bibfnamefont {Vedran}\ \bibnamefont
  {Dunjko}}\ and\ \bibinfo {author} {\bibfnamefont {Hans~J}\ \bibnamefont
  {Briegel}},\ }\bibfield  {title} {\enquote {\bibinfo {title} {Machine
  learning {\&} artificial intelligence in the quantum domain: a review of
  recent progress},}\ }\href {\doibase 10.1088/1361-6633/aab406} {\bibfield
  {journal} {\bibinfo  {journal} {Reports on Progress in Physics}\ }\textbf
  {\bibinfo {volume} {81}},\ \bibinfo {pages} {074001} (\bibinfo {year}
  {2018})}\BibitemShut {NoStop}%
\bibitem [{\citenamefont {Mehta}\ \emph {et~al.}(2019)\citenamefont {Mehta},
  \citenamefont {Bukov}, \citenamefont {Wang}, \citenamefont {Day},
  \citenamefont {Richardson}, \citenamefont {Fisher},\ and\ \citenamefont
  {Schwab}}]{Mehta_2019}%
  \BibitemOpen
  \bibfield  {author} {\bibinfo {author} {\bibfnamefont {Pankaj}\ \bibnamefont
  {Mehta}}, \bibinfo {author} {\bibfnamefont {Marin}\ \bibnamefont {Bukov}},
  \bibinfo {author} {\bibfnamefont {Ching-Hao}\ \bibnamefont {Wang}}, \bibinfo
  {author} {\bibfnamefont {Alexandre~G.R.}\ \bibnamefont {Day}}, \bibinfo
  {author} {\bibfnamefont {Clint}\ \bibnamefont {Richardson}}, \bibinfo
  {author} {\bibfnamefont {Charles~K.}\ \bibnamefont {Fisher}}, \ and\ \bibinfo
  {author} {\bibfnamefont {David~J.}\ \bibnamefont {Schwab}},\ }\bibfield
  {title} {\enquote {\bibinfo {title} {A high-bias, low-variance introduction
  to machine learning for physicists},}\ }\href {\doibase
  https://doi.org/10.1016/j.physrep.2019.03.001} {\bibfield  {journal}
  {\bibinfo  {journal} {Physics Reports}\ }\textbf {\bibinfo {volume} {810}},\
  \bibinfo {pages} {1 -- 124} (\bibinfo {year} {2019})},\ \bibinfo {note} {a
  high-bias, low-variance introduction to Machine Learning for
  physicists}\BibitemShut {NoStop}%
\bibitem [{\citenamefont {Carleo}\ \emph {et~al.}(2019)\citenamefont {Carleo},
  \citenamefont {Cirac}, \citenamefont {Cranmer}, \citenamefont {Daudet},
  \citenamefont {Schuld}, \citenamefont {Tishby}, \citenamefont
  {Vogt-Maranto},\ and\ \citenamefont {Zdeborov\'a}}]{Carleo_2019}%
  \BibitemOpen
  \bibfield  {author} {\bibinfo {author} {\bibfnamefont {Giuseppe}\
  \bibnamefont {Carleo}}, \bibinfo {author} {\bibfnamefont {Ignacio}\
  \bibnamefont {Cirac}}, \bibinfo {author} {\bibfnamefont {Kyle}\ \bibnamefont
  {Cranmer}}, \bibinfo {author} {\bibfnamefont {Laurent}\ \bibnamefont
  {Daudet}}, \bibinfo {author} {\bibfnamefont {Maria}\ \bibnamefont {Schuld}},
  \bibinfo {author} {\bibfnamefont {Naftali}\ \bibnamefont {Tishby}}, \bibinfo
  {author} {\bibfnamefont {Leslie}\ \bibnamefont {Vogt-Maranto}}, \ and\
  \bibinfo {author} {\bibfnamefont {Lenka}\ \bibnamefont {Zdeborov\'a}},\
  }\bibfield  {title} {\enquote {\bibinfo {title} {Machine learning and the
  physical sciences},}\ }\href {\doibase 10.1103/RevModPhys.91.045002}
  {\bibfield  {journal} {\bibinfo  {journal} {Rev. Mod. Phys.}\ }\textbf
  {\bibinfo {volume} {91}},\ \bibinfo {pages} {045002} (\bibinfo {year}
  {2019})}\BibitemShut {NoStop}%
\bibitem [{\citenamefont {Ghiringhelli}\ \emph {et~al.}(2015)\citenamefont
  {Ghiringhelli}, \citenamefont {Vybiral}, \citenamefont {Levchenko},
  \citenamefont {Draxl},\ and\ \citenamefont {Scheffler}}]{Ghiringhelli2015}%
  \BibitemOpen
  \bibfield  {author} {\bibinfo {author} {\bibfnamefont {Luca~M.}\ \bibnamefont
  {Ghiringhelli}}, \bibinfo {author} {\bibfnamefont {Jan}\ \bibnamefont
  {Vybiral}}, \bibinfo {author} {\bibfnamefont {Sergey~V.}\ \bibnamefont
  {Levchenko}}, \bibinfo {author} {\bibfnamefont {Claudia}\ \bibnamefont
  {Draxl}}, \ and\ \bibinfo {author} {\bibfnamefont {Matthias}\ \bibnamefont
  {Scheffler}},\ }\bibfield  {title} {\enquote {\bibinfo {title} {Big data of
  materials science: Critical role of the descriptor},}\ }\href {\doibase
  10.1103/PhysRevLett.114.105503} {\bibfield  {journal} {\bibinfo  {journal}
  {Phys. Rev. Lett.}\ }\textbf {\bibinfo {volume} {114}},\ \bibinfo {pages}
  {105503} (\bibinfo {year} {2015})}\BibitemShut {NoStop}%
\bibitem [{\citenamefont {Torlai}\ and\ \citenamefont
  {Melko}(2016)}]{Torlai2016}%
  \BibitemOpen
  \bibfield  {author} {\bibinfo {author} {\bibfnamefont {Giacomo}\ \bibnamefont
  {Torlai}}\ and\ \bibinfo {author} {\bibfnamefont {Roger~G.}\ \bibnamefont
  {Melko}},\ }\bibfield  {title} {\enquote {\bibinfo {title} {Learning
  thermodynamics with boltzmann machines},}\ }\href {\doibase
  10.1103/PhysRevB.94.165134} {\bibfield  {journal} {\bibinfo  {journal} {Phys.
  Rev. B}\ }\textbf {\bibinfo {volume} {94}},\ \bibinfo {pages} {165134}
  (\bibinfo {year} {2016})}\BibitemShut {NoStop}%
\bibitem [{\citenamefont {Carleo}\ and\ \citenamefont
  {Troyer}(2017)}]{Carleo2017}%
  \BibitemOpen
  \bibfield  {author} {\bibinfo {author} {\bibfnamefont {Giuseppe}\
  \bibnamefont {Carleo}}\ and\ \bibinfo {author} {\bibfnamefont {Matthias}\
  \bibnamefont {Troyer}},\ }\bibfield  {title} {\enquote {\bibinfo {title}
  {Solving the quantum many-body problem with artificial neural networks},}\
  }\href {\doibase 10.1126/science.aag2302} {\bibfield  {journal} {\bibinfo
  {journal} {Science}\ }\textbf {\bibinfo {volume} {355}},\ \bibinfo {pages}
  {602--606} (\bibinfo {year} {2017})}\BibitemShut {NoStop}%
\bibitem [{\citenamefont {Carrasquilla}\ and\ \citenamefont
  {Melko}(2017)}]{Carrasquilla_2017}%
  \BibitemOpen
  \bibfield  {author} {\bibinfo {author} {\bibfnamefont {Juan}\ \bibnamefont
  {Carrasquilla}}\ and\ \bibinfo {author} {\bibfnamefont {Roger~G.}\
  \bibnamefont {Melko}},\ }\bibfield  {title} {\enquote {\bibinfo {title}
  {Machine learning phases of matter},}\ }\href {\doibase 10.1038/nphys4035}
  {\bibfield  {journal} {\bibinfo  {journal} {Nature Physics}\ }\textbf
  {\bibinfo {volume} {13}},\ \bibinfo {pages} {431--434} (\bibinfo {year}
  {2017})}\BibitemShut {NoStop}%
\bibitem [{\citenamefont {Ponte}\ and\ \citenamefont
  {Melko}(2017)}]{Ponte_2017}%
  \BibitemOpen
  \bibfield  {author} {\bibinfo {author} {\bibfnamefont {Pedro}\ \bibnamefont
  {Ponte}}\ and\ \bibinfo {author} {\bibfnamefont {Roger~G.}\ \bibnamefont
  {Melko}},\ }\bibfield  {title} {\enquote {\bibinfo {title} {Kernel methods
  for interpretable machine learning of order parameters},}\ }\href@noop {}
  {\bibfield  {journal} {\bibinfo  {journal} {Phys. Rev. B}\ }\textbf {\bibinfo
  {volume} {96}},\ \bibinfo {pages} {205146} (\bibinfo {year}
  {2017})}\BibitemShut {NoStop}%
\bibitem [{\citenamefont {Liu}\ \emph {et~al.}(2019)\citenamefont {Liu},
  \citenamefont {Greitemann},\ and\ \citenamefont {Pollet}}]{Liu_2019}%
  \BibitemOpen
  \bibfield  {author} {\bibinfo {author} {\bibfnamefont {Ke}~\bibnamefont
  {Liu}}, \bibinfo {author} {\bibfnamefont {Jonas}\ \bibnamefont {Greitemann}},
  \ and\ \bibinfo {author} {\bibfnamefont {Lode}\ \bibnamefont {Pollet}},\
  }\bibfield  {title} {\enquote {\bibinfo {title} {Learning multiple order
  parameters with interpretable machines},}\ }\href {\doibase
  10.1103/PhysRevB.99.104410} {\bibfield  {journal} {\bibinfo  {journal} {Phys.
  Rev. B}\ }\textbf {\bibinfo {volume} {99}},\ \bibinfo {pages} {104410}
  (\bibinfo {year} {2019})}\BibitemShut {NoStop}%
\bibitem [{\citenamefont {Canabarro}\ \emph
  {et~al.}(2019{\natexlab{a}})\citenamefont {Canabarro}, \citenamefont
  {Fanchini}, \citenamefont {Malvezzi}, \citenamefont {Pereira},\ and\
  \citenamefont {Chaves}}]{Canabarro_2019}%
  \BibitemOpen
  \bibfield  {author} {\bibinfo {author} {\bibfnamefont {Askery}\ \bibnamefont
  {Canabarro}}, \bibinfo {author} {\bibfnamefont {Felipe~Fernandes}\
  \bibnamefont {Fanchini}}, \bibinfo {author} {\bibfnamefont {Andr\'e~Luiz}\
  \bibnamefont {Malvezzi}}, \bibinfo {author} {\bibfnamefont {Rodrigo}\
  \bibnamefont {Pereira}}, \ and\ \bibinfo {author} {\bibfnamefont {Rafael}\
  \bibnamefont {Chaves}},\ }\bibfield  {title} {\enquote {\bibinfo {title}
  {Unveiling phase transitions with machine learning},}\ }\href {\doibase
  10.1103/PhysRevB.100.045129} {\bibfield  {journal} {\bibinfo  {journal}
  {Phys. Rev. B}\ }\textbf {\bibinfo {volume} {100}},\ \bibinfo {pages}
  {045129} (\bibinfo {year} {2019}{\natexlab{a}})}\BibitemShut {NoStop}%
\bibitem [{\citenamefont {Torlai}\ \emph {et~al.}(2018)\citenamefont {Torlai},
  \citenamefont {Mazzola}, \citenamefont {Carrasquilla}, \citenamefont
  {Troyer}, \citenamefont {Melko},\ and\ \citenamefont {Carleo}}]{Torlai2018}%
  \BibitemOpen
  \bibfield  {author} {\bibinfo {author} {\bibfnamefont {Giacomo}\ \bibnamefont
  {Torlai}}, \bibinfo {author} {\bibfnamefont {Guglielmo}\ \bibnamefont
  {Mazzola}}, \bibinfo {author} {\bibfnamefont {Juan}\ \bibnamefont
  {Carrasquilla}}, \bibinfo {author} {\bibfnamefont {Matthias}\ \bibnamefont
  {Troyer}}, \bibinfo {author} {\bibfnamefont {Roger}\ \bibnamefont {Melko}}, \
  and\ \bibinfo {author} {\bibfnamefont {Giuseppe}\ \bibnamefont {Carleo}},\
  }\bibfield  {title} {\enquote {\bibinfo {title} {Neural-network quantum state
  tomography},}\ }\href {\doibase 10.1038/s41567-018-0048-5} {\bibfield
  {journal} {\bibinfo  {journal} {Nature Physics}\ }\textbf {\bibinfo {volume}
  {14}},\ \bibinfo {pages} {447--450} (\bibinfo {year} {2018})}\BibitemShut
  {NoStop}%
\bibitem [{\citenamefont {Canabarro}\ \emph
  {et~al.}(2019{\natexlab{b}})\citenamefont {Canabarro}, \citenamefont
  {Brito},\ and\ \citenamefont {Chaves}}]{Canabarro2019}%
  \BibitemOpen
  \bibfield  {author} {\bibinfo {author} {\bibfnamefont {Askery}\ \bibnamefont
  {Canabarro}}, \bibinfo {author} {\bibfnamefont {Samura\'{\i}}\ \bibnamefont
  {Brito}}, \ and\ \bibinfo {author} {\bibfnamefont {Rafael}\ \bibnamefont
  {Chaves}},\ }\bibfield  {title} {\enquote {\bibinfo {title} {Machine learning
  nonlocal correlations},}\ }\href {\doibase 10.1103/PhysRevLett.122.200401}
  {\bibfield  {journal} {\bibinfo  {journal} {Phys. Rev. Lett.}\ }\textbf
  {\bibinfo {volume} {122}},\ \bibinfo {pages} {200401} (\bibinfo {year}
  {2019}{\natexlab{b}})}\BibitemShut {NoStop}%
\bibitem [{\citenamefont {Iten}\ \emph {et~al.}(2020)\citenamefont {Iten},
  \citenamefont {Metger}, \citenamefont {Wilming}, \citenamefont {del Rio},\
  and\ \citenamefont {Renner}}]{Raban2020}%
  \BibitemOpen
  \bibfield  {author} {\bibinfo {author} {\bibfnamefont {Raban}\ \bibnamefont
  {Iten}}, \bibinfo {author} {\bibfnamefont {Tony}\ \bibnamefont {Metger}},
  \bibinfo {author} {\bibfnamefont {Henrik}\ \bibnamefont {Wilming}}, \bibinfo
  {author} {\bibfnamefont {L\'{\i}dia}\ \bibnamefont {del Rio}}, \ and\
  \bibinfo {author} {\bibfnamefont {Renato}\ \bibnamefont {Renner}},\
  }\bibfield  {title} {\enquote {\bibinfo {title} {Discovering physical
  concepts with neural networks},}\ }\href {\doibase
  10.1103/PhysRevLett.124.010508} {\bibfield  {journal} {\bibinfo  {journal}
  {Phys. Rev. Lett.}\ }\textbf {\bibinfo {volume} {124}},\ \bibinfo {pages}
  {010508} (\bibinfo {year} {2020})}\BibitemShut {NoStop}%
\bibitem [{\citenamefont {Breuer}\ and\ \citenamefont
  {Petruccione}(2007)}]{BreuerPet}%
  \BibitemOpen
  \bibfield  {author} {\bibinfo {author} {\bibfnamefont {H.{-}P.}\ \bibnamefont
  {Breuer}}\ and\ \bibinfo {author} {\bibfnamefont {F.}~\bibnamefont
  {Petruccione}},\ }\href@noop {} {\emph {\bibinfo {title} {The {T}heory of
  {O}pen {Q}uantum {S}ystems}}}\ (\bibinfo  {publisher} {Oxford University
  Press, Oxford},\ \bibinfo {year} {2007})\BibitemShut {NoStop}%
\bibitem [{\citenamefont {Rivas}\ and\ \citenamefont
  {Huelga}(2012)}]{Rivas2012}%
  \BibitemOpen
  \bibfield  {author} {\bibinfo {author} {\bibfnamefont {Angel}\ \bibnamefont
  {Rivas}}\ and\ \bibinfo {author} {\bibfnamefont {Susana~F}\ \bibnamefont
  {Huelga}},\ }\href@noop {} {\emph {\bibinfo {title} {Open quantum systems}}}\
  (\bibinfo  {publisher} {Springer},\ \bibinfo {year} {2012})\BibitemShut
  {NoStop}%
\bibitem [{\citenamefont {Baumgratz}\ \emph {et~al.}(2014)\citenamefont
  {Baumgratz}, \citenamefont {Cramer},\ and\ \citenamefont
  {Plenio}}]{Baumgratz2014}%
  \BibitemOpen
  \bibfield  {author} {\bibinfo {author} {\bibfnamefont {T.}~\bibnamefont
  {Baumgratz}}, \bibinfo {author} {\bibfnamefont {M.}~\bibnamefont {Cramer}}, \
  and\ \bibinfo {author} {\bibfnamefont {M.~B.}\ \bibnamefont {Plenio}},\
  }\bibfield  {title} {\enquote {\bibinfo {title} {Quantifying coherence},}\
  }\href {\doibase 10.1103/PhysRevLett.113.140401} {\bibfield  {journal}
  {\bibinfo  {journal} {Phys. Rev. Lett.}\ }\textbf {\bibinfo {volume} {113}},\
  \bibinfo {pages} {140401} (\bibinfo {year} {2014})}\BibitemShut {NoStop}%
\bibitem [{\citenamefont {Streltsov}\ \emph {et~al.}(2017)\citenamefont
  {Streltsov}, \citenamefont {Adesso},\ and\ \citenamefont
  {Plenio}}]{Streltsov2017}%
  \BibitemOpen
  \bibfield  {author} {\bibinfo {author} {\bibfnamefont {Alexander}\
  \bibnamefont {Streltsov}}, \bibinfo {author} {\bibfnamefont {Gerardo}\
  \bibnamefont {Adesso}}, \ and\ \bibinfo {author} {\bibfnamefont {Martin~B.}\
  \bibnamefont {Plenio}},\ }\bibfield  {title} {\enquote {\bibinfo {title}
  {Colloquium: Quantum coherence as a resource},}\ }\href {\doibase
  10.1103/RevModPhys.89.041003} {\bibfield  {journal} {\bibinfo  {journal}
  {Rev. Mod. Phys.}\ }\textbf {\bibinfo {volume} {89}},\ \bibinfo {pages}
  {041003} (\bibinfo {year} {2017})}\BibitemShut {NoStop}%
\bibitem [{\citenamefont {Breuer}\ \emph {et~al.}(2016)\citenamefont {Breuer},
  \citenamefont {Laine}, \citenamefont {Piilo},\ and\ \citenamefont
  {Vacchini}}]{Breuer2016}%
  \BibitemOpen
  \bibfield  {author} {\bibinfo {author} {\bibfnamefont {Heinz-Peter}\
  \bibnamefont {Breuer}}, \bibinfo {author} {\bibfnamefont {Elsi-Mari}\
  \bibnamefont {Laine}}, \bibinfo {author} {\bibfnamefont {Jyrki}\ \bibnamefont
  {Piilo}}, \ and\ \bibinfo {author} {\bibfnamefont {Bassano}\ \bibnamefont
  {Vacchini}},\ }\bibfield  {title} {\enquote {\bibinfo {title} {Colloquium:
  Non-markovian dynamics in open quantum systems},}\ }\href {\doibase
  10.1103/RevModPhys.88.021002} {\bibfield  {journal} {\bibinfo  {journal}
  {Rev. Mod. Phys.}\ }\textbf {\bibinfo {volume} {88}},\ \bibinfo {pages}
  {021002} (\bibinfo {year} {2016})}\BibitemShut {NoStop}%
\bibitem [{\citenamefont {Li}\ \emph {et~al.}(2018)\citenamefont {Li},
  \citenamefont {Hall},\ and\ \citenamefont {Wiseman}}]{Li2018}%
  \BibitemOpen
  \bibfield  {author} {\bibinfo {author} {\bibfnamefont {Li}~\bibnamefont
  {Li}}, \bibinfo {author} {\bibfnamefont {Michael~J.W.}\ \bibnamefont {Hall}},
  \ and\ \bibinfo {author} {\bibfnamefont {Howard~M.}\ \bibnamefont
  {Wiseman}},\ }\bibfield  {title} {\enquote {\bibinfo {title} {Concepts of
  quantum non-markovianity: A hierarchy},}\ }\href {\doibase
  https://doi.org/10.1016/j.physrep.2018.07.001} {\bibfield  {journal}
  {\bibinfo  {journal} {Physics Reports}\ }\textbf {\bibinfo {volume} {759}},\
  \bibinfo {pages} {1 -- 51} (\bibinfo {year} {2018})}\BibitemShut {NoStop}%
\bibitem [{\citenamefont {Li}\ \emph {et~al.}(2020{\natexlab{a}})\citenamefont
  {Li}, \citenamefont {Guo},\ and\ \citenamefont {Piilo}}]{Li2020x}%
  \BibitemOpen
  \bibfield  {author} {\bibinfo {author} {\bibfnamefont {C.-F.}\ \bibnamefont
  {Li}}, \bibinfo {author} {\bibfnamefont {G.-C.}\ \bibnamefont {Guo}}, \ and\
  \bibinfo {author} {\bibfnamefont {J.}~\bibnamefont {Piilo}},\ }\bibfield
  {title} {\enquote {\bibinfo {title} {Non-markovian quantum dynamics: What is
  it good for?}}\ }\href {\doibase 10.1209/0295-5075/128/30001} {\bibfield
  {journal} {\bibinfo  {journal} {Europhys. Lett.}\ }\textbf {\bibinfo {volume}
  {128}},\ \bibinfo {pages} {30001} (\bibinfo {year}
  {2020}{\natexlab{a}})}\BibitemShut {NoStop}%
\bibitem [{\citenamefont {Fanchini}\ \emph {et~al.}(2013)\citenamefont
  {Fanchini}, \citenamefont {Karpat}, \citenamefont {Castelano},\ and\
  \citenamefont {Rossatto}}]{Fanchini2013}%
  \BibitemOpen
  \bibfield  {author} {\bibinfo {author} {\bibfnamefont {Felipe~F.}\
  \bibnamefont {Fanchini}}, \bibinfo {author} {\bibfnamefont {G\"oktu\u{g}}\
  \bibnamefont {Karpat}}, \bibinfo {author} {\bibfnamefont {Leonardo~K.}\
  \bibnamefont {Castelano}}, \ and\ \bibinfo {author} {\bibfnamefont
  {Daniel~Z.}\ \bibnamefont {Rossatto}},\ }\bibfield  {title} {\enquote
  {\bibinfo {title} {Probing the degree of non-markovianity for independent and
  common environments},}\ }\href {\doibase 10.1103/PhysRevA.88.012105}
  {\bibfield  {journal} {\bibinfo  {journal} {Phys. Rev. A}\ }\textbf {\bibinfo
  {volume} {88}},\ \bibinfo {pages} {012105} (\bibinfo {year}
  {2013})}\BibitemShut {NoStop}%
\bibitem [{\citenamefont {Addis}\ \emph {et~al.}(2016)\citenamefont {Addis},
  \citenamefont {Karpat}, \citenamefont {Macchiavello},\ and\ \citenamefont
  {Maniscalco}}]{Addis2016}%
  \BibitemOpen
  \bibfield  {author} {\bibinfo {author} {\bibfnamefont {Carole}\ \bibnamefont
  {Addis}}, \bibinfo {author} {\bibfnamefont {G\"oktu\u{g}}\ \bibnamefont
  {Karpat}}, \bibinfo {author} {\bibfnamefont {Chiara}\ \bibnamefont
  {Macchiavello}}, \ and\ \bibinfo {author} {\bibfnamefont {Sabrina}\
  \bibnamefont {Maniscalco}},\ }\bibfield  {title} {\enquote {\bibinfo {title}
  {Dynamical memory effects in correlated quantum channels},}\ }\href {\doibase
  10.1103/PhysRevA.94.032121} {\bibfield  {journal} {\bibinfo  {journal} {Phys.
  Rev. A}\ }\textbf {\bibinfo {volume} {94}},\ \bibinfo {pages} {032121}
  (\bibinfo {year} {2016})}\BibitemShut {NoStop}%
\bibitem [{\citenamefont {Rivas}\ \emph {et~al.}(2014)\citenamefont {Rivas},
  \citenamefont {Huelga},\ and\ \citenamefont {Plenio}}]{Rivas2014}%
  \BibitemOpen
  \bibfield  {author} {\bibinfo {author} {\bibfnamefont {{\'{A}}ngel}\
  \bibnamefont {Rivas}}, \bibinfo {author} {\bibfnamefont {Susana~F}\
  \bibnamefont {Huelga}}, \ and\ \bibinfo {author} {\bibfnamefont {Martin~B}\
  \bibnamefont {Plenio}},\ }\bibfield  {title} {\enquote {\bibinfo {title}
  {Quantum non-markovianity: characterization, quantification and detection},}\
  }\href {\doibase 10.1088/0034-4885/77/9/094001} {\bibfield  {journal}
  {\bibinfo  {journal} {Rep. Prog. Phys.}\ }\textbf {\bibinfo {volume} {77}},\
  \bibinfo {pages} {094001} (\bibinfo {year} {2014})}\BibitemShut {NoStop}%
\bibitem [{\citenamefont {Liu}\ \emph {et~al.}(2011)\citenamefont {Liu},
  \citenamefont {Li}, \citenamefont {Huang}, \citenamefont {Li}, \citenamefont
  {Guo}, \citenamefont {Laine}, \citenamefont {Breuer},\ and\ \citenamefont
  {Piilo}}]{Liu2011}%
  \BibitemOpen
  \bibfield  {author} {\bibinfo {author} {\bibfnamefont {Bi-Heng}\ \bibnamefont
  {Liu}}, \bibinfo {author} {\bibfnamefont {Li}~\bibnamefont {Li}}, \bibinfo
  {author} {\bibfnamefont {Yun-Feng}\ \bibnamefont {Huang}}, \bibinfo {author}
  {\bibfnamefont {Chuan-Feng}\ \bibnamefont {Li}}, \bibinfo {author}
  {\bibfnamefont {Guang-Can}\ \bibnamefont {Guo}}, \bibinfo {author}
  {\bibfnamefont {Elsi-Mari}\ \bibnamefont {Laine}}, \bibinfo {author}
  {\bibfnamefont {Heinz-Peter}\ \bibnamefont {Breuer}}, \ and\ \bibinfo
  {author} {\bibfnamefont {Jyrki}\ \bibnamefont {Piilo}},\ }\bibfield  {title}
  {\enquote {\bibinfo {title} {Experimental control of the transition from
  markovian to non-markovian dynamics of open quantum systems},}\ }\href
  {\doibase 10.1038/nphys2085} {\bibfield  {journal} {\bibinfo  {journal} {Nat.
  Phys.}\ }\textbf {\bibinfo {volume} {7}},\ \bibinfo {pages} {931--934}
  (\bibinfo {year} {2011})}\BibitemShut {NoStop}%
\bibitem [{\citenamefont {Fanchini}\ \emph {et~al.}(2014)\citenamefont
  {Fanchini}, \citenamefont {Karpat}, \citenamefont {\ifmmode~\mbox{\c{C}}\else
  \c{C}\fi{}akmak}, \citenamefont {Castelano}, \citenamefont {Aguilar},
  \citenamefont {Far\'{\i}as}, \citenamefont {Walborn}, \citenamefont
  {Ribeiro},\ and\ \citenamefont {de~Oliveira}}]{Fanchini2014}%
  \BibitemOpen
  \bibfield  {author} {\bibinfo {author} {\bibfnamefont {F.~F.}\ \bibnamefont
  {Fanchini}}, \bibinfo {author} {\bibfnamefont {G.}~\bibnamefont {Karpat}},
  \bibinfo {author} {\bibfnamefont {B.}~\bibnamefont
  {\ifmmode~\mbox{\c{C}}\else \c{C}\fi{}akmak}}, \bibinfo {author}
  {\bibfnamefont {L.~K.}\ \bibnamefont {Castelano}}, \bibinfo {author}
  {\bibfnamefont {G.~H.}\ \bibnamefont {Aguilar}}, \bibinfo {author}
  {\bibfnamefont {O.~Jim\'enez}\ \bibnamefont {Far\'{\i}as}}, \bibinfo {author}
  {\bibfnamefont {S.~P.}\ \bibnamefont {Walborn}}, \bibinfo {author}
  {\bibfnamefont {P.~H.~Souto}\ \bibnamefont {Ribeiro}}, \ and\ \bibinfo
  {author} {\bibfnamefont {M.~C.}\ \bibnamefont {de~Oliveira}},\ }\bibfield
  {title} {\enquote {\bibinfo {title} {Non-markovianity through accessible
  information},}\ }\href {\doibase 10.1103/PhysRevLett.112.210402} {\bibfield
  {journal} {\bibinfo  {journal} {Phys. Rev. Lett.}\ }\textbf {\bibinfo
  {volume} {112}},\ \bibinfo {pages} {210402} (\bibinfo {year}
  {2014})}\BibitemShut {NoStop}%
\bibitem [{\citenamefont {Haseli}\ \emph {et~al.}(2014)\citenamefont {Haseli},
  \citenamefont {Karpat}, \citenamefont {Salimi}, \citenamefont {Khorashad},
  \citenamefont {Fanchini}, \citenamefont {\ifmmode~\mbox{\c{C}}\else
  \c{C}\fi{}akmak}, \citenamefont {Aguilar}, \citenamefont {Walborn},\ and\
  \citenamefont {Ribeiro}}]{Haseli2014}%
  \BibitemOpen
  \bibfield  {author} {\bibinfo {author} {\bibfnamefont {S.}~\bibnamefont
  {Haseli}}, \bibinfo {author} {\bibfnamefont {G.}~\bibnamefont {Karpat}},
  \bibinfo {author} {\bibfnamefont {S.}~\bibnamefont {Salimi}}, \bibinfo
  {author} {\bibfnamefont {A.~S.}\ \bibnamefont {Khorashad}}, \bibinfo {author}
  {\bibfnamefont {F.~F.}\ \bibnamefont {Fanchini}}, \bibinfo {author}
  {\bibfnamefont {B.}~\bibnamefont {\ifmmode~\mbox{\c{C}}\else
  \c{C}\fi{}akmak}}, \bibinfo {author} {\bibfnamefont {G.~H.}\ \bibnamefont
  {Aguilar}}, \bibinfo {author} {\bibfnamefont {S.~P.}\ \bibnamefont
  {Walborn}}, \ and\ \bibinfo {author} {\bibfnamefont {P.~H.~Souto}\
  \bibnamefont {Ribeiro}},\ }\bibfield  {title} {\enquote {\bibinfo {title}
  {Non-markovianity through flow of information between a system and an
  environment},}\ }\href {\doibase 10.1103/PhysRevA.90.052118} {\bibfield
  {journal} {\bibinfo  {journal} {Phys. Rev. A}\ }\textbf {\bibinfo {volume}
  {90}},\ \bibinfo {pages} {052118} (\bibinfo {year} {2014})}\BibitemShut
  {NoStop}%
\bibitem [{\citenamefont {Li}\ \emph {et~al.}(2020{\natexlab{b}})\citenamefont
  {Li}, \citenamefont {Guo},\ and\ \citenamefont {Piilo}}]{Li2020}%
  \BibitemOpen
  \bibfield  {author} {\bibinfo {author} {\bibfnamefont {C.-F.}\ \bibnamefont
  {Li}}, \bibinfo {author} {\bibfnamefont {G.-C.}\ \bibnamefont {Guo}}, \ and\
  \bibinfo {author} {\bibfnamefont {J.}~\bibnamefont {Piilo}},\ }\bibfield
  {title} {\enquote {\bibinfo {title} {Non-markovian quantum dynamics: What is
  it good for?}}\ }\href {\doibase 10.1209/0295-5075/128/30001} {\bibfield
  {journal} {\bibinfo  {journal} {Europhys. Lett.}\ }\textbf {\bibinfo {volume}
  {128}},\ \bibinfo {pages} {30001} (\bibinfo {year}
  {2020}{\natexlab{b}})}\BibitemShut {NoStop}%
\bibitem [{\citenamefont {Banchi}\ \emph {et~al.}(2018)\citenamefont {Banchi},
  \citenamefont {Grant}, \citenamefont {Rocchetto},\ and\ \citenamefont
  {Severini}}]{Banchi2018}%
  \BibitemOpen
  \bibfield  {author} {\bibinfo {author} {\bibfnamefont {Leonardo}\
  \bibnamefont {Banchi}}, \bibinfo {author} {\bibfnamefont {Edward}\
  \bibnamefont {Grant}}, \bibinfo {author} {\bibfnamefont {Andrea}\
  \bibnamefont {Rocchetto}}, \ and\ \bibinfo {author} {\bibfnamefont {Simone}\
  \bibnamefont {Severini}},\ }\bibfield  {title} {\enquote {\bibinfo {title}
  {Modelling non-markovian quantum processes with recurrent neural networks},}\
  }\href {\doibase 10.1088/1367-2630/aaf749} {\bibfield  {journal} {\bibinfo
  {journal} {New Journal of Physics}\ }\textbf {\bibinfo {volume} {20}},\
  \bibinfo {pages} {123030} (\bibinfo {year} {2018})}\BibitemShut {NoStop}%
\bibitem [{\citenamefont {Shrapnel}\ \emph {et~al.}(2018)\citenamefont
  {Shrapnel}, \citenamefont {Costa},\ and\ \citenamefont
  {Milburn}}]{Shrapnel2018}%
  \BibitemOpen
  \bibfield  {author} {\bibinfo {author} {\bibfnamefont {Sally}\ \bibnamefont
  {Shrapnel}}, \bibinfo {author} {\bibfnamefont {Fabio}\ \bibnamefont {Costa}},
  \ and\ \bibinfo {author} {\bibfnamefont {Gerard}\ \bibnamefont {Milburn}},\
  }\bibfield  {title} {\enquote {\bibinfo {title} {Quantum markovianity as a
  supervised learning task},}\ }\href {\doibase 10.1142/S0219749918400105}
  {\bibfield  {journal} {\bibinfo  {journal} {International Journal of Quantum
  Information}\ }\textbf {\bibinfo {volume} {16}},\ \bibinfo {pages} {1840010}
  (\bibinfo {year} {2018})}\BibitemShut {NoStop}%
\bibitem [{\citenamefont {Luchnikov}\ \emph {et~al.}(2020)\citenamefont
  {Luchnikov}, \citenamefont {Vintskevich}, \citenamefont {Grigoriev},\ and\
  \citenamefont {Filippov}}]{Luchnikov2020}%
  \BibitemOpen
  \bibfield  {author} {\bibinfo {author} {\bibfnamefont {I.~A.}\ \bibnamefont
  {Luchnikov}}, \bibinfo {author} {\bibfnamefont {S.~V.}\ \bibnamefont
  {Vintskevich}}, \bibinfo {author} {\bibfnamefont {D.~A.}\ \bibnamefont
  {Grigoriev}}, \ and\ \bibinfo {author} {\bibfnamefont {S.~N.}\ \bibnamefont
  {Filippov}},\ }\bibfield  {title} {\enquote {\bibinfo {title} {Machine
  learning non-markovian quantum dynamics},}\ }\href {\doibase
  10.1103/PhysRevLett.124.140502} {\bibfield  {journal} {\bibinfo  {journal}
  {Phys. Rev. Lett.}\ }\textbf {\bibinfo {volume} {124}},\ \bibinfo {pages}
  {140502} (\bibinfo {year} {2020})}\BibitemShut {NoStop}%
\bibitem [{\citenamefont {Chu~Guo}(2020)}]{2004.11038}%
  \BibitemOpen
  \bibfield  {author} {\bibinfo {author} {\bibfnamefont {Dario~Poletti}\
  \bibnamefont {Chu~Guo}, \bibfnamefont {Kavan~Modi}},\ }\href@noop {}
  {\enquote {\bibinfo {title} {Tensor network based machine learning of
  non-markovian quantum processes},}\ } (\bibinfo {year} {2020}),\ \Eprint
  {http://arxiv.org/abs/arXiv:2004.11038} {arXiv:2004.11038} \BibitemShut
  {NoStop}%
\bibitem [{\citenamefont {Breuer}\ \emph {et~al.}(2009)\citenamefont {Breuer},
  \citenamefont {Laine},\ and\ \citenamefont {Piilo}}]{Breuer2009}%
  \BibitemOpen
  \bibfield  {author} {\bibinfo {author} {\bibfnamefont {Heinz-Peter}\
  \bibnamefont {Breuer}}, \bibinfo {author} {\bibfnamefont {Elsi-Mari}\
  \bibnamefont {Laine}}, \ and\ \bibinfo {author} {\bibfnamefont {Jyrki}\
  \bibnamefont {Piilo}},\ }\bibfield  {title} {\enquote {\bibinfo {title}
  {Measure for the degree of non-markovian behavior of quantum processes in
  open systems},}\ }\href {\doibase 10.1103/PhysRevLett.103.210401} {\bibfield
  {journal} {\bibinfo  {journal} {Phys. Rev. Lett.}\ }\textbf {\bibinfo
  {volume} {103}},\ \bibinfo {pages} {210401} (\bibinfo {year}
  {2009})}\BibitemShut {NoStop}%
\bibitem [{\citenamefont {Rivas}\ \emph {et~al.}(2010)\citenamefont {Rivas},
  \citenamefont {Huelga},\ and\ \citenamefont {Plenio}}]{Rivas2010}%
  \BibitemOpen
  \bibfield  {author} {\bibinfo {author} {\bibfnamefont {\'Angel}\ \bibnamefont
  {Rivas}}, \bibinfo {author} {\bibfnamefont {Susana~F.}\ \bibnamefont
  {Huelga}}, \ and\ \bibinfo {author} {\bibfnamefont {Martin~B.}\ \bibnamefont
  {Plenio}},\ }\bibfield  {title} {\enquote {\bibinfo {title} {Entanglement and
  non-markovianity of quantum evolutions},}\ }\href {\doibase
  10.1103/PhysRevLett.105.050403} {\bibfield  {journal} {\bibinfo  {journal}
  {Phys. Rev. Lett.}\ }\textbf {\bibinfo {volume} {105}},\ \bibinfo {pages}
  {050403} (\bibinfo {year} {2010})}\BibitemShut {NoStop}%
\bibitem [{\citenamefont {Laine}\ \emph {et~al.}(2014)\citenamefont {Laine},
  \citenamefont {Breuer},\ and\ \citenamefont {Piilo}}]{Laine2014}%
  \BibitemOpen
  \bibfield  {author} {\bibinfo {author} {\bibfnamefont {Elsi-Mari}\
  \bibnamefont {Laine}}, \bibinfo {author} {\bibfnamefont {Heinz-Peter}\
  \bibnamefont {Breuer}}, \ and\ \bibinfo {author} {\bibfnamefont {Jyrki}\
  \bibnamefont {Piilo}},\ }\bibfield  {title} {\enquote {\bibinfo {title}
  {Nonlocal memory effects allow perfect teleportation with mixed states},}\
  }\href {\doibase 10.1038/srep04620} {\bibfield  {journal} {\bibinfo
  {journal} {Scientific Reports}\ }\textbf {\bibinfo {volume} {4}},\ \bibinfo
  {pages} {4620} (\bibinfo {year} {2014})}\BibitemShut {NoStop}%
\bibitem [{\citenamefont {Liu}\ \emph {et~al.}(2016)\citenamefont {Liu},
  \citenamefont {Hu}, \citenamefont {Huang}, \citenamefont {Li}, \citenamefont
  {Guo}, \citenamefont {Karlsson}, \citenamefont {Laine}, \citenamefont
  {Maniscalco}, \citenamefont {Macchiavello},\ and\ \citenamefont
  {Piilo}}]{Liu2016}%
  \BibitemOpen
  \bibfield  {author} {\bibinfo {author} {\bibfnamefont {Bi-Heng}\ \bibnamefont
  {Liu}}, \bibinfo {author} {\bibfnamefont {Xiao-Min}\ \bibnamefont {Hu}},
  \bibinfo {author} {\bibfnamefont {Yun-Feng}\ \bibnamefont {Huang}}, \bibinfo
  {author} {\bibfnamefont {Chuan-Feng}\ \bibnamefont {Li}}, \bibinfo {author}
  {\bibfnamefont {Guang-Can}\ \bibnamefont {Guo}}, \bibinfo {author}
  {\bibfnamefont {Antti}\ \bibnamefont {Karlsson}}, \bibinfo {author}
  {\bibfnamefont {Elsi-Mari}\ \bibnamefont {Laine}}, \bibinfo {author}
  {\bibfnamefont {Sabrina}\ \bibnamefont {Maniscalco}}, \bibinfo {author}
  {\bibfnamefont {Chiara}\ \bibnamefont {Macchiavello}}, \ and\ \bibinfo
  {author} {\bibfnamefont {Jyrki}\ \bibnamefont {Piilo}},\ }\bibfield  {title}
  {\enquote {\bibinfo {title} {Efficient superdense coding in the presence of
  non-markovian noise},}\ }\href {\doibase 10.1209/0295-5075/114/10005}
  {\bibfield  {journal} {\bibinfo  {journal} {{EPL} (Europhysics Letters)}\
  }\textbf {\bibinfo {volume} {114}},\ \bibinfo {pages} {10005} (\bibinfo
  {year} {2016})}\BibitemShut {NoStop}%
\bibitem [{\citenamefont {Karpat}\ \emph {et~al.}(2015)\citenamefont {Karpat},
  \citenamefont {Piilo},\ and\ \citenamefont {Maniscalco}}]{Karpat2015}%
  \BibitemOpen
  \bibfield  {author} {\bibinfo {author} {\bibfnamefont {Göktu{\u{g}}}\
  \bibnamefont {Karpat}}, \bibinfo {author} {\bibfnamefont {Jyrki}\
  \bibnamefont {Piilo}}, \ and\ \bibinfo {author} {\bibfnamefont {Sabrina}\
  \bibnamefont {Maniscalco}},\ }\bibfield  {title} {\enquote {\bibinfo {title}
  {Controlling entropic uncertainty bound through memory effects},}\ }\href
  {\doibase 10.1209/0295-5075/111/50006} {\bibfield  {journal} {\bibinfo
  {journal} {{EPL} (Europhysics Letters)}\ }\textbf {\bibinfo {volume} {111}},\
  \bibinfo {pages} {50006} (\bibinfo {year} {2015})}\BibitemShut {NoStop}%
\bibitem [{\citenamefont {Berta}\ \emph {et~al.}(2010)\citenamefont {Berta},
  \citenamefont {Christandl}, \citenamefont {Colbeck}, \citenamefont {Renes},\
  and\ \citenamefont {Renner}}]{Berta2010}%
  \BibitemOpen
  \bibfield  {author} {\bibinfo {author} {\bibfnamefont {Mario}\ \bibnamefont
  {Berta}}, \bibinfo {author} {\bibfnamefont {Matthias}\ \bibnamefont
  {Christandl}}, \bibinfo {author} {\bibfnamefont {Roger}\ \bibnamefont
  {Colbeck}}, \bibinfo {author} {\bibfnamefont {Joseph~M.}\ \bibnamefont
  {Renes}}, \ and\ \bibinfo {author} {\bibfnamefont {Renato}\ \bibnamefont
  {Renner}},\ }\bibfield  {title} {\enquote {\bibinfo {title} {The uncertainty
  principle in the presence of quantum memory},}\ }\href {\doibase
  10.1038/nphys1734} {\bibfield  {journal} {\bibinfo  {journal} {Nature
  Physics}\ }\textbf {\bibinfo {volume} {6}},\ \bibinfo {pages} {659--662}
  (\bibinfo {year} {2010})}\BibitemShut {NoStop}%
\bibitem [{\citenamefont {Bylicka}\ \emph {et~al.}(2016)\citenamefont
  {Bylicka}, \citenamefont {Tukiainen}, \citenamefont
  {Chru{\'{s}}ci{\'{n}}ski}, \citenamefont {Piilo},\ and\ \citenamefont
  {Maniscalco}}]{Bylicka2016}%
  \BibitemOpen
  \bibfield  {author} {\bibinfo {author} {\bibfnamefont {Bogna}\ \bibnamefont
  {Bylicka}}, \bibinfo {author} {\bibfnamefont {Mikko}\ \bibnamefont
  {Tukiainen}}, \bibinfo {author} {\bibfnamefont {Dariusz}\ \bibnamefont
  {Chru{\'{s}}ci{\'{n}}ski}}, \bibinfo {author} {\bibfnamefont {Jyrki}\
  \bibnamefont {Piilo}}, \ and\ \bibinfo {author} {\bibfnamefont {Sabrina}\
  \bibnamefont {Maniscalco}},\ }\bibfield  {title} {\enquote {\bibinfo {title}
  {Thermodynamic power of non-markovianity},}\ }\href {\doibase
  10.1038/srep27989} {\bibfield  {journal} {\bibinfo  {journal} {Scientific
  Reports}\ }\textbf {\bibinfo {volume} {6}},\ \bibinfo {pages} {27989}
  (\bibinfo {year} {2016})}\BibitemShut {NoStop}%
\bibitem [{\citenamefont {Bylicka}\ \emph {et~al.}(2014)\citenamefont
  {Bylicka}, \citenamefont {Chru{\'{s}}ci{\'{n}}ski},\ and\ \citenamefont
  {Maniscalco}}]{Bylicka2014}%
  \BibitemOpen
  \bibfield  {author} {\bibinfo {author} {\bibfnamefont {B.}~\bibnamefont
  {Bylicka}}, \bibinfo {author} {\bibfnamefont {D.}~\bibnamefont
  {Chru{\'{s}}ci{\'{n}}ski}}, \ and\ \bibinfo {author} {\bibfnamefont
  {S.}~\bibnamefont {Maniscalco}},\ }\bibfield  {title} {\enquote {\bibinfo
  {title} {Non-markovianity and reservoir memory of quantum channels: a quantum
  information theory perspective},}\ }\href {\doibase 10.1038/srep05720}
  {\bibfield  {journal} {\bibinfo  {journal} {Scientific Reports}\ }\textbf
  {\bibinfo {volume} {4}},\ \bibinfo {pages} {5720} (\bibinfo {year}
  {2014})}\BibitemShut {NoStop}%
\bibitem [{\citenamefont {Giorgi}\ \emph {et~al.}(2012)\citenamefont {Giorgi},
  \citenamefont {Galve}, \citenamefont {Manzano}, \citenamefont {Colet},\ and\
  \citenamefont {Zambrini}}]{giorgi2012}%
  \BibitemOpen
  \bibfield  {author} {\bibinfo {author} {\bibfnamefont {Gian~Luca}\
  \bibnamefont {Giorgi}}, \bibinfo {author} {\bibfnamefont {Fernando}\
  \bibnamefont {Galve}}, \bibinfo {author} {\bibfnamefont {Gonzalo}\
  \bibnamefont {Manzano}}, \bibinfo {author} {\bibfnamefont {Pere}\
  \bibnamefont {Colet}}, \ and\ \bibinfo {author} {\bibfnamefont {Roberta}\
  \bibnamefont {Zambrini}},\ }\bibfield  {title} {\enquote {\bibinfo {title}
  {Quantum correlations and mutual synchronization},}\ }\href {\doibase
  10.1103/PhysRevA.85.052101} {\bibfield  {journal} {\bibinfo  {journal} {Phys.
  Rev. A}\ }\textbf {\bibinfo {volume} {85}},\ \bibinfo {pages} {052101}
  (\bibinfo {year} {2012})}\BibitemShut {NoStop}%
\bibitem [{\citenamefont {Karpat}\ \emph {et~al.}(2020)\citenamefont {Karpat},
  \citenamefont {İskender Yalçınkaya}, \citenamefont {Çakmak},
  \citenamefont {Giorgi},\ and\ \citenamefont {Zambrini}}]{Karpat2020}%
  \BibitemOpen
  \bibfield  {author} {\bibinfo {author} {\bibfnamefont {Göktuğ}\
  \bibnamefont {Karpat}}, \bibinfo {author} {\bibnamefont {İskender
  Yalçınkaya}}, \bibinfo {author} {\bibfnamefont {Barış}\ \bibnamefont
  {Çakmak}}, \bibinfo {author} {\bibfnamefont {Gian~Luca}\ \bibnamefont
  {Giorgi}}, \ and\ \bibinfo {author} {\bibfnamefont {Roberta}\ \bibnamefont
  {Zambrini}},\ }\href@noop {} {\enquote {\bibinfo {title} {Synchronization and
  non-markovianity in open quantum systems},}\ } (\bibinfo {year} {2020}),\
  \Eprint {http://arxiv.org/abs/2008.03310} {arXiv:2008.03310 [quant-ph]}
  \BibitemShut {NoStop}%
\bibitem [{\citenamefont {Vacchini}\ \emph {et~al.}(2011)\citenamefont
  {Vacchini}, \citenamefont {Smirne}, \citenamefont {Laine}, \citenamefont
  {Piilo},\ and\ \citenamefont {Breuer}}]{Vacchini2011}%
  \BibitemOpen
  \bibfield  {author} {\bibinfo {author} {\bibfnamefont {Bassano}\ \bibnamefont
  {Vacchini}}, \bibinfo {author} {\bibfnamefont {Andrea}\ \bibnamefont
  {Smirne}}, \bibinfo {author} {\bibfnamefont {Elsi-Mari}\ \bibnamefont
  {Laine}}, \bibinfo {author} {\bibfnamefont {Jyrki}\ \bibnamefont {Piilo}}, \
  and\ \bibinfo {author} {\bibfnamefont {Heinz-Peter}\ \bibnamefont {Breuer}},\
  }\bibfield  {title} {\enquote {\bibinfo {title} {Markovianity and
  non-markovianity in quantum and classical systems},}\ }\href {\doibase
  10.1088/1367-2630/13/9/093004} {\bibfield  {journal} {\bibinfo  {journal}
  {New Journal of Physics}\ }\textbf {\bibinfo {volume} {13}},\ \bibinfo
  {pages} {093004} (\bibinfo {year} {2011})}\BibitemShut {NoStop}%
\bibitem [{\citenamefont {Lindblad}(1976)}]{Lindblad1976}%
  \BibitemOpen
  \bibfield  {author} {\bibinfo {author} {\bibfnamefont {G.}~\bibnamefont
  {Lindblad}},\ }\bibfield  {title} {\enquote {\bibinfo {title} {On the
  generators of quantum dynamical semigroups},}\ }\href {\doibase
  10.1007/BF01608499} {\bibfield  {journal} {\bibinfo  {journal}
  {Communications in Mathematical Physics}\ }\textbf {\bibinfo {volume} {48}},\
  \bibinfo {pages} {119--130} (\bibinfo {year} {1976})}\BibitemShut {NoStop}%
\bibitem [{\citenamefont {Gorini}\ \emph {et~al.}(1976)\citenamefont {Gorini},
  \citenamefont {Kossakowski},\ and\ \citenamefont {Sudarshan}}]{gorini1976}%
  \BibitemOpen
  \bibfield  {author} {\bibinfo {author} {\bibfnamefont {Vittorio}\
  \bibnamefont {Gorini}}, \bibinfo {author} {\bibfnamefont {Andrzej}\
  \bibnamefont {Kossakowski}}, \ and\ \bibinfo {author} {\bibfnamefont
  {E.~C.~G.}\ \bibnamefont {Sudarshan}},\ }\bibfield  {title} {\enquote
  {\bibinfo {title} {Completely positive dynamical semigroups of n‐level
  systems},}\ }\href {\doibase 10.1063/1.522979} {\bibfield  {journal}
  {\bibinfo  {journal} {Journal of Mathematical Physics}\ }\textbf {\bibinfo
  {volume} {17}},\ \bibinfo {pages} {821--825} (\bibinfo {year} {1976})},\
  \Eprint
  {http://arxiv.org/abs/https://aip.scitation.org/doi/pdf/10.1063/1.522979}
  {https://aip.scitation.org/doi/pdf/10.1063/1.522979} \BibitemShut {NoStop}%
\bibitem [{\citenamefont {Breuer}(2012)}]{Breuer2012}%
  \BibitemOpen
  \bibfield  {author} {\bibinfo {author} {\bibfnamefont {Heinz-Peter}\
  \bibnamefont {Breuer}},\ }\bibfield  {title} {\enquote {\bibinfo {title}
  {Foundations and measures of quantum non-markovianity},}\ }\href {\doibase
  10.1088/0953-4075/45/15/154001} {\bibfield  {journal} {\bibinfo  {journal}
  {Journal of Physics B: Atomic, Molecular and Optical Physics}\ }\textbf
  {\bibinfo {volume} {45}},\ \bibinfo {pages} {154001} (\bibinfo {year}
  {2012})}\BibitemShut {NoStop}%
\bibitem [{\citenamefont {Addis}\ \emph {et~al.}(2014)\citenamefont {Addis},
  \citenamefont {Bylicka}, \citenamefont {Chru\ifmmode \acute{s}\else
  \'{s}\fi{}ci\ifmmode~\acute{n}\else \'{n}\fi{}ski},\ and\ \citenamefont
  {Maniscalco}}]{Addis2014}%
  \BibitemOpen
  \bibfield  {author} {\bibinfo {author} {\bibfnamefont {Carole}\ \bibnamefont
  {Addis}}, \bibinfo {author} {\bibfnamefont {Bogna}\ \bibnamefont {Bylicka}},
  \bibinfo {author} {\bibfnamefont {Dariusz}\ \bibnamefont {Chru\ifmmode
  \acute{s}\else \'{s}\fi{}ci\ifmmode~\acute{n}\else \'{n}\fi{}ski}}, \ and\
  \bibinfo {author} {\bibfnamefont {Sabrina}\ \bibnamefont {Maniscalco}},\
  }\bibfield  {title} {\enquote {\bibinfo {title} {Comparative study of
  non-markovianity measures in exactly solvable one- and two-qubit models},}\
  }\href {\doibase 10.1103/PhysRevA.90.052103} {\bibfield  {journal} {\bibinfo
  {journal} {Phys. Rev. A}\ }\textbf {\bibinfo {volume} {90}},\ \bibinfo
  {pages} {052103} (\bibinfo {year} {2014})}\BibitemShut {NoStop}%
\bibitem [{\citenamefont {Wi\ss{}mann}\ \emph {et~al.}(2012)\citenamefont
  {Wi\ss{}mann}, \citenamefont {Karlsson}, \citenamefont {Laine}, \citenamefont
  {Piilo},\ and\ \citenamefont {Breuer}}]{Wismannn2012}%
  \BibitemOpen
  \bibfield  {author} {\bibinfo {author} {\bibfnamefont {Steffen}\ \bibnamefont
  {Wi\ss{}mann}}, \bibinfo {author} {\bibfnamefont {Antti}\ \bibnamefont
  {Karlsson}}, \bibinfo {author} {\bibfnamefont {Elsi-Mari}\ \bibnamefont
  {Laine}}, \bibinfo {author} {\bibfnamefont {Jyrki}\ \bibnamefont {Piilo}}, \
  and\ \bibinfo {author} {\bibfnamefont {Heinz-Peter}\ \bibnamefont {Breuer}},\
  }\bibfield  {title} {\enquote {\bibinfo {title} {Optimal state pairs for
  non-markovian quantum dynamics},}\ }\href {\doibase
  10.1103/PhysRevA.86.062108} {\bibfield  {journal} {\bibinfo  {journal} {Phys.
  Rev. A}\ }\textbf {\bibinfo {volume} {86}},\ \bibinfo {pages} {062108}
  (\bibinfo {year} {2012})}\BibitemShut {NoStop}%
\bibitem [{\citenamefont {Neto}\ \emph {et~al.}(2016)\citenamefont {Neto},
  \citenamefont {Karpat},\ and\ \citenamefont {Fanchini}}]{Neto2016}%
  \BibitemOpen
  \bibfield  {author} {\bibinfo {author} {\bibfnamefont {Alaor~Cervati}\
  \bibnamefont {Neto}}, \bibinfo {author} {\bibfnamefont {G\"oktu\u{g}}\
  \bibnamefont {Karpat}}, \ and\ \bibinfo {author} {\bibfnamefont
  {Felipe~Fernandes}\ \bibnamefont {Fanchini}},\ }\bibfield  {title} {\enquote
  {\bibinfo {title} {Inequivalence of correlation-based measures of
  non-markovianity},}\ }\href {\doibase 10.1103/PhysRevA.94.032105} {\bibfield
  {journal} {\bibinfo  {journal} {Phys. Rev. A}\ }\textbf {\bibinfo {volume}
  {94}},\ \bibinfo {pages} {032105} (\bibinfo {year} {2016})}\BibitemShut
  {NoStop}%
\bibitem [{\citenamefont {Hill}\ and\ \citenamefont
  {Wootters}(1997)}]{Wootters1997}%
  \BibitemOpen
  \bibfield  {author} {\bibinfo {author} {\bibfnamefont {Scott}\ \bibnamefont
  {Hill}}\ and\ \bibinfo {author} {\bibfnamefont {William~K.}\ \bibnamefont
  {Wootters}},\ }\bibfield  {title} {\enquote {\bibinfo {title} {Entanglement
  of a pair of quantum bits},}\ }\href {\doibase 10.1103/PhysRevLett.78.5022}
  {\bibfield  {journal} {\bibinfo  {journal} {Phys. Rev. Lett.}\ }\textbf
  {\bibinfo {volume} {78}},\ \bibinfo {pages} {5022--5025} (\bibinfo {year}
  {1997})}\BibitemShut {NoStop}%
\bibitem [{\citenamefont {Daffer}\ \emph {et~al.}(2004)\citenamefont {Daffer},
  \citenamefont {W\'odkiewicz}, \citenamefont {Cresser},\ and\ \citenamefont
  {McIver}}]{daffer04}%
  \BibitemOpen
  \bibfield  {author} {\bibinfo {author} {\bibfnamefont {Sonja}\ \bibnamefont
  {Daffer}}, \bibinfo {author} {\bibfnamefont {Krzysztof}\ \bibnamefont
  {W\'odkiewicz}}, \bibinfo {author} {\bibfnamefont {James~D.}\ \bibnamefont
  {Cresser}}, \ and\ \bibinfo {author} {\bibfnamefont {John~K.}\ \bibnamefont
  {McIver}},\ }\bibfield  {title} {\enquote {\bibinfo {title} {Depolarizing
  channel as a completely positive map with memory},}\ }\href {\doibase
  10.1103/PhysRevA.70.010304} {\bibfield  {journal} {\bibinfo  {journal} {Phys.
  Rev. A}\ }\textbf {\bibinfo {volume} {70}},\ \bibinfo {pages} {010304}
  (\bibinfo {year} {2004})}\BibitemShut {NoStop}%
\bibitem [{\citenamefont {Whalen}\ and\ \citenamefont
  {Carmichael}(2016)}]{whalen2016}%
  \BibitemOpen
  \bibfield  {author} {\bibinfo {author} {\bibfnamefont {S.~J.}\ \bibnamefont
  {Whalen}}\ and\ \bibinfo {author} {\bibfnamefont {H.~J.}\ \bibnamefont
  {Carmichael}},\ }\bibfield  {title} {\enquote {\bibinfo {title} {Time-local
  {H}eisenberg-{L}angevin equations and the driven qubit},}\ }\href {\doibase
  10.1103/PhysRevA.93.063820} {\bibfield  {journal} {\bibinfo  {journal} {Phys.
  Rev. A}\ }\textbf {\bibinfo {volume} {93}},\ \bibinfo {pages} {063820}
  (\bibinfo {year} {2016})}\BibitemShut {NoStop}%
\bibitem [{\citenamefont {Haikka}\ and\ \citenamefont
  {Maniscalco}(2010)}]{Haikka2010}%
  \BibitemOpen
  \bibfield  {author} {\bibinfo {author} {\bibfnamefont {P.}~\bibnamefont
  {Haikka}}\ and\ \bibinfo {author} {\bibfnamefont {S.}~\bibnamefont
  {Maniscalco}},\ }\bibfield  {title} {\enquote {\bibinfo {title}
  {Non-markovian dynamics of a damped driven two-state system},}\ }\href
  {\doibase 10.1103/PhysRevA.81.052103} {\bibfield  {journal} {\bibinfo
  {journal} {Phys. Rev. A}\ }\textbf {\bibinfo {volume} {81}},\ \bibinfo
  {pages} {052103} (\bibinfo {year} {2010})}\BibitemShut {NoStop}%
\bibitem [{\citenamefont {Haikka}(2010)}]{Haikka2010-2}%
  \BibitemOpen
  \bibfield  {author} {\bibinfo {author} {\bibfnamefont {P}~\bibnamefont
  {Haikka}},\ }\bibfield  {title} {\enquote {\bibinfo {title} {Non-{M}arkovian
  master equation for a damped driven two-state system},}\ }\href {\doibase
  10.1088/0031-8949/2010/t140/014047} {\bibfield  {journal} {\bibinfo
  {journal} {Phys. Scr.}\ }\textbf {\bibinfo {volume} {T140}},\ \bibinfo
  {pages} {014047} (\bibinfo {year} {2010})}\BibitemShut {NoStop}%
\bibitem [{\citenamefont {Shen}\ \emph {et~al.}(2014)\citenamefont {Shen},
  \citenamefont {Qin}, \citenamefont {Xiu},\ and\ \citenamefont
  {Yi}}]{Shen2014}%
  \BibitemOpen
  \bibfield  {author} {\bibinfo {author} {\bibfnamefont {H.~Z.}\ \bibnamefont
  {Shen}}, \bibinfo {author} {\bibfnamefont {M.}~\bibnamefont {Qin}}, \bibinfo
  {author} {\bibfnamefont {Xiao-Ming}\ \bibnamefont {Xiu}}, \ and\ \bibinfo
  {author} {\bibfnamefont {X.~X.}\ \bibnamefont {Yi}},\ }\bibfield  {title}
  {\enquote {\bibinfo {title} {Exact non-{M}arkovian master equation for a
  driven damped two-level system},}\ }\href {\doibase
  10.1103/PhysRevA.89.062113} {\bibfield  {journal} {\bibinfo  {journal} {Phys.
  Rev. A}\ }\textbf {\bibinfo {volume} {89}},\ \bibinfo {pages} {062113}
  (\bibinfo {year} {2014})}\BibitemShut {NoStop}%
\bibitem [{\citenamefont {Huang}\ and\ \citenamefont {Situ}(2017)}]{Huang2017}%
  \BibitemOpen
  \bibfield  {author} {\bibinfo {author} {\bibfnamefont {Zhiming}\ \bibnamefont
  {Huang}}\ and\ \bibinfo {author} {\bibfnamefont {Haozhen}\ \bibnamefont
  {Situ}},\ }\bibfield  {title} {\enquote {\bibinfo {title} {Non-markovian
  dynamics of quantum coherence of two-level system driven by classical
  field},}\ }\href {\doibase 10.1007/s11128-017-1673-0} {\bibfield  {journal}
  {\bibinfo  {journal} {Quantum Inf. Process.}\ }\textbf {\bibinfo {volume}
  {16}},\ \bibinfo {pages} {222} (\bibinfo {year} {2017})}\BibitemShut
  {NoStop}%
\bibitem [{\citenamefont {Bellomo}\ \emph {et~al.}(2007)\citenamefont
  {Bellomo}, \citenamefont {Lo~Franco},\ and\ \citenamefont
  {Compagno}}]{Bellomo2007}%
  \BibitemOpen
  \bibfield  {author} {\bibinfo {author} {\bibfnamefont {B.}~\bibnamefont
  {Bellomo}}, \bibinfo {author} {\bibfnamefont {R.}~\bibnamefont {Lo~Franco}},
  \ and\ \bibinfo {author} {\bibfnamefont {G.}~\bibnamefont {Compagno}},\
  }\bibfield  {title} {\enquote {\bibinfo {title} {{Non-Markovian Effects on
  the Dynamics of Entanglement}},}\ }\href {\doibase
  10.1103/PhysRevLett.99.160502} {\bibfield  {journal} {\bibinfo  {journal}
  {Phys. Rev. Lett.}\ }\textbf {\bibinfo {volume} {99}},\ \bibinfo {pages}
  {160502} (\bibinfo {year} {2007})}\BibitemShut {NoStop}%
\bibitem [{\citenamefont {Garraway}(1997)}]{Garraway1997}%
  \BibitemOpen
  \bibfield  {author} {\bibinfo {author} {\bibfnamefont {B.~M.}\ \bibnamefont
  {Garraway}},\ }\bibfield  {title} {\enquote {\bibinfo {title}
  {Nonperturbative decay of an atomic system in a cavity},}\ }\href {\doibase
  10.1103/PhysRevA.55.2290} {\bibfield  {journal} {\bibinfo  {journal} {Phys.
  Rev. A}\ }\textbf {\bibinfo {volume} {55}},\ \bibinfo {pages} {2290--2303}
  (\bibinfo {year} {1997})}\BibitemShut {NoStop}%
\bibitem [{\citenamefont {Vapnik}(1995)}]{Vapnik_1995}%
  \BibitemOpen
  \bibfield  {author} {\bibinfo {author} {\bibfnamefont {V.}~\bibnamefont
  {Vapnik}},\ }\bibfield  {title} {\enquote {\bibinfo {title} {The nature of
  statistical learning theory.}}\ }\href@noop {} {\bibfield  {journal}
  {\bibinfo  {journal} {Springer Verlag, New York}\ } (\bibinfo {year}
  {1995})}\BibitemShut {NoStop}%
\bibitem [{\citenamefont {Burges}(1998)}]{Burges_1998}%
  \BibitemOpen
  \bibfield  {author} {\bibinfo {author} {\bibfnamefont {Christopher J.~C.}\
  \bibnamefont {Burges}},\ }\bibfield  {title} {\enquote {\bibinfo {title} {A
  tutorial on support vector machines for pattern recognition},}\ }\href
  {\doibase 10.1023/A:1009715923555} {\bibfield  {journal} {\bibinfo  {journal}
  {Data Mining and Knowledge Discovery}\ }\textbf {\bibinfo {volume} {2}},\
  \bibinfo {pages} {121--167} (\bibinfo {year} {1998})}\BibitemShut {NoStop}%
\bibitem [{\citenamefont {Dietrich}\ \emph {et~al.}(1999)\citenamefont
  {Dietrich}, \citenamefont {Opper},\ and\ \citenamefont
  {Sompolinsky}}]{Dietrich_1999}%
  \BibitemOpen
  \bibfield  {author} {\bibinfo {author} {\bibfnamefont {Rainer}\ \bibnamefont
  {Dietrich}}, \bibinfo {author} {\bibfnamefont {Manfred}\ \bibnamefont
  {Opper}}, \ and\ \bibinfo {author} {\bibfnamefont {Haim}\ \bibnamefont
  {Sompolinsky}},\ }\bibfield  {title} {\enquote {\bibinfo {title} {Statistical
  mechanics of support vector networks},}\ }\href {\doibase
  10.1103/PhysRevLett.82.2975} {\bibfield  {journal} {\bibinfo  {journal}
  {Phys. Rev. Lett.}\ }\textbf {\bibinfo {volume} {82}},\ \bibinfo {pages}
  {2975--2978} (\bibinfo {year} {1999})}\BibitemShut {NoStop}%
\bibitem [{\citenamefont {Risau-Gusman}\ and\ \citenamefont
  {Gordon}(2000)}]{Risau_2000}%
  \BibitemOpen
  \bibfield  {author} {\bibinfo {author} {\bibfnamefont {Sebastian}\
  \bibnamefont {Risau-Gusman}}\ and\ \bibinfo {author} {\bibfnamefont
  {Mirta~B.}\ \bibnamefont {Gordon}},\ }\bibfield  {title} {\enquote {\bibinfo
  {title} {Generalization properties of finite-size polynomial support vector
  machines},}\ }\href {\doibase 10.1103/PhysRevE.62.7092} {\bibfield  {journal}
  {\bibinfo  {journal} {Phys. Rev. E}\ }\textbf {\bibinfo {volume} {62}},\
  \bibinfo {pages} {7092--7099} (\bibinfo {year} {2000})}\BibitemShut {NoStop}%
\bibitem [{\citenamefont {Opper}\ and\ \citenamefont
  {Urbanczik}(2001)}]{Opper_2001}%
  \BibitemOpen
  \bibfield  {author} {\bibinfo {author} {\bibfnamefont {M.}~\bibnamefont
  {Opper}}\ and\ \bibinfo {author} {\bibfnamefont {R.}~\bibnamefont
  {Urbanczik}},\ }\bibfield  {title} {\enquote {\bibinfo {title} {Universal
  learning curves of support vector machines},}\ }\href {\doibase
  10.1103/PhysRevLett.86.4410} {\bibfield  {journal} {\bibinfo  {journal}
  {Phys. Rev. Lett.}\ }\textbf {\bibinfo {volume} {86}},\ \bibinfo {pages}
  {4410--4413} (\bibinfo {year} {2001})}\BibitemShut {NoStop}%
\bibitem [{\citenamefont {B.}\ \emph {et~al.}(1998)\citenamefont {B.},
  \citenamefont {P.},\ and\ \citenamefont {Smola~A.}}]{Scholkopf_1998}%
  \BibitemOpen
  \bibfield  {author} {\bibinfo {author} {\bibfnamefont {Sch{\"o}lkopf}\
  \bibnamefont {B.}}, \bibinfo {author} {\bibfnamefont {Bartlett}\ \bibnamefont
  {P.}}, \ and\ \bibinfo {author} {\bibfnamefont {Williamson~R.}\ \bibnamefont
  {Smola~A.}},\ }\bibfield  {title} {\enquote {\bibinfo {title} {Support vector
  regression with automatic accuracy control.}}\ }\href@noop {} {\bibfield
  {journal} {\bibinfo  {journal} {In: Niklasson L., Bod{\'e}n M., Ziemke T.
  (eds) ICANN 98. ICANN 1998. Perspectives in Neural Computing. Springer,
  London}\ } (\bibinfo {year} {1998})}\BibitemShut {NoStop}%
\bibitem [{\citenamefont {B.}\ and\ \citenamefont
  {A.J.}(2002)}]{Scholkpf_2002}%
  \BibitemOpen
  \bibfield  {author} {\bibinfo {author} {\bibfnamefont {Sch{\"o}lkopf}\
  \bibnamefont {B.}}\ and\ \bibinfo {author} {\bibfnamefont {Smola}\
  \bibnamefont {A.J.}},\ }\href@noop {} {\emph {\bibinfo {title} {Learning with
  Kernels}}}\ (\bibinfo  {publisher} {MIT Press.},\ \bibinfo {year}
  {2002})\BibitemShut {NoStop}%
\bibitem [{\citenamefont {Smola}\ and\ \citenamefont
  {Sch{\"o}lkopf}(2004)}]{Smola_2004}%
  \BibitemOpen
  \bibfield  {author} {\bibinfo {author} {\bibfnamefont {Alex~J.}\ \bibnamefont
  {Smola}}\ and\ \bibinfo {author} {\bibfnamefont {Bernhard}\ \bibnamefont
  {Sch{\"o}lkopf}},\ }\bibfield  {title} {\enquote {\bibinfo {title} {A
  tutorial on support vector regression},}\ }\href {\doibase
  10.1023/B:STCO.0000035301.49549.88} {\bibfield  {journal} {\bibinfo
  {journal} {Statistics and Computing}\ }\textbf {\bibinfo {volume} {14}},\
  \bibinfo {pages} {199--222} (\bibinfo {year} {2004})}\BibitemShut {NoStop}%
\bibitem [{\citenamefont {Drucker}\ \emph
  {et~al.}(1996{\natexlab{a}})\citenamefont {Drucker}, \citenamefont {Burges},
  \citenamefont {Kaufman}, \citenamefont {Smola},\ and\ \citenamefont
  {Vapnik}}]{Drucker96}%
  \BibitemOpen
  \bibfield  {author} {\bibinfo {author} {\bibfnamefont {Harris}\ \bibnamefont
  {Drucker}}, \bibinfo {author} {\bibfnamefont {Chris J.~C.}\ \bibnamefont
  {Burges}}, \bibinfo {author} {\bibfnamefont {Linda}\ \bibnamefont {Kaufman}},
  \bibinfo {author} {\bibfnamefont {Alex}\ \bibnamefont {Smola}}, \ and\
  \bibinfo {author} {\bibfnamefont {Vladimir}\ \bibnamefont {Vapnik}},\
  }\bibfield  {title} {\enquote {\bibinfo {title} {Support vector regression
  machines},}\ \ }(\bibinfo  {publisher} {MIT Press},\ \bibinfo {address}
  {Cambridge, MA, USA},\ \bibinfo {year} {1996})\ p.\ \bibinfo {pages}
  {155–161}\BibitemShut {NoStop}%
\bibitem [{\citenamefont {Rebentrost}\ \emph {et~al.}(2014)\citenamefont
  {Rebentrost}, \citenamefont {Mohseni},\ and\ \citenamefont
  {Lloyd}}]{Rebentrost_2014}%
  \BibitemOpen
  \bibfield  {author} {\bibinfo {author} {\bibfnamefont {Patrick}\ \bibnamefont
  {Rebentrost}}, \bibinfo {author} {\bibfnamefont {Masoud}\ \bibnamefont
  {Mohseni}}, \ and\ \bibinfo {author} {\bibfnamefont {Seth}\ \bibnamefont
  {Lloyd}},\ }\bibfield  {title} {\enquote {\bibinfo {title} {Quantum support
  vector machine for big data classification},}\ }\href {\doibase
  10.1103/PhysRevLett.113.130503} {\bibfield  {journal} {\bibinfo  {journal}
  {Phys. Rev. Lett.}\ }\textbf {\bibinfo {volume} {113}},\ \bibinfo {pages}
  {130503} (\bibinfo {year} {2014})}\BibitemShut {NoStop}%
\bibitem [{\citenamefont {Li}\ \emph {et~al.}(2015)\citenamefont {Li},
  \citenamefont {Liu}, \citenamefont {Xu},\ and\ \citenamefont {Du}}]{Li_2015}%
  \BibitemOpen
  \bibfield  {author} {\bibinfo {author} {\bibfnamefont {Zhaokai}\ \bibnamefont
  {Li}}, \bibinfo {author} {\bibfnamefont {Xiaomei}\ \bibnamefont {Liu}},
  \bibinfo {author} {\bibfnamefont {Nanyang}\ \bibnamefont {Xu}}, \ and\
  \bibinfo {author} {\bibfnamefont {Jiangfeng}\ \bibnamefont {Du}},\ }\bibfield
   {title} {\enquote {\bibinfo {title} {Experimental realization of a quantum
  support vector machine},}\ }\href {\doibase 10.1103/PhysRevLett.114.140504}
  {\bibfield  {journal} {\bibinfo  {journal} {Phys. Rev. Lett.}\ }\textbf
  {\bibinfo {volume} {114}},\ \bibinfo {pages} {140504} (\bibinfo {year}
  {2015})}\BibitemShut {NoStop}%
\bibitem [{\citenamefont {Biamonte}\ \emph {et~al.}(2017)\citenamefont
  {Biamonte}, \citenamefont {Wittek}, \citenamefont {Pancotti}, \citenamefont
  {Rebentrost}, \citenamefont {Wiebe},\ and\ \citenamefont
  {Lloyd}}]{Biamonte_2017}%
  \BibitemOpen
  \bibfield  {author} {\bibinfo {author} {\bibfnamefont {Jacob}\ \bibnamefont
  {Biamonte}}, \bibinfo {author} {\bibfnamefont {Peter}\ \bibnamefont
  {Wittek}}, \bibinfo {author} {\bibfnamefont {Nicola}\ \bibnamefont
  {Pancotti}}, \bibinfo {author} {\bibfnamefont {Patrick}\ \bibnamefont
  {Rebentrost}}, \bibinfo {author} {\bibfnamefont {Nathan}\ \bibnamefont
  {Wiebe}}, \ and\ \bibinfo {author} {\bibfnamefont {Seth}\ \bibnamefont
  {Lloyd}},\ }\bibfield  {title} {\enquote {\bibinfo {title} {Quantum machine
  learning},}\ }\href {\doibase 10.1038/nature23474} {\bibfield  {journal}
  {\bibinfo  {journal} {Nature}\ }\textbf {\bibinfo {volume} {549}},\ \bibinfo
  {pages} {195--202} (\bibinfo {year} {2017})}\BibitemShut {NoStop}%
\bibitem [{\citenamefont {Havl{\'\i}{\v c}ek}\ \emph
  {et~al.}(2019)\citenamefont {Havl{\'\i}{\v c}ek}, \citenamefont
  {C{\'o}rcoles}, \citenamefont {Temme}, \citenamefont {Harrow}, \citenamefont
  {Kandala}, \citenamefont {Chow},\ and\ \citenamefont
  {Gambetta}}]{Havlicek_2019}%
  \BibitemOpen
  \bibfield  {author} {\bibinfo {author} {\bibfnamefont {Vojt{\v e}ch}\
  \bibnamefont {Havl{\'\i}{\v c}ek}}, \bibinfo {author} {\bibfnamefont
  {Antonio~D.}\ \bibnamefont {C{\'o}rcoles}}, \bibinfo {author} {\bibfnamefont
  {Kristan}\ \bibnamefont {Temme}}, \bibinfo {author} {\bibfnamefont {Aram~W.}\
  \bibnamefont {Harrow}}, \bibinfo {author} {\bibfnamefont {Abhinav}\
  \bibnamefont {Kandala}}, \bibinfo {author} {\bibfnamefont {Jerry~M.}\
  \bibnamefont {Chow}}, \ and\ \bibinfo {author} {\bibfnamefont {Jay~M.}\
  \bibnamefont {Gambetta}},\ }\bibfield  {title} {\enquote {\bibinfo {title}
  {Supervised learning with quantum-enhanced feature spaces},}\ }\href
  {\doibase 10.1038/s41586-019-0980-2} {\bibfield  {journal} {\bibinfo
  {journal} {Nature}\ }\textbf {\bibinfo {volume} {567}},\ \bibinfo {pages}
  {209--212} (\bibinfo {year} {2019})}\BibitemShut {NoStop}%
\bibitem [{\citenamefont {Ioffe}\ and\ \citenamefont
  {Szegedy}(2015)}]{ioffe2015}%
  \BibitemOpen
  \bibfield  {author} {\bibinfo {author} {\bibfnamefont {Sergey}\ \bibnamefont
  {Ioffe}}\ and\ \bibinfo {author} {\bibfnamefont {Christian}\ \bibnamefont
  {Szegedy}},\ }\href@noop {} {\enquote {\bibinfo {title} {Batch normalization:
  Accelerating deep network training by reducing internal covariate shift},}\ }
  (\bibinfo {year} {2015}),\ \Eprint {http://arxiv.org/abs/1502.03167}
  {arXiv:1502.03167} \BibitemShut {NoStop}%
\bibitem [{\citenamefont {Wise}\ \emph {et~al.}(2020)\citenamefont {Wise},
  \citenamefont {Morton},\ and\ \citenamefont {Dhomkar}}]{2005.01144}%
  \BibitemOpen
  \bibfield  {author} {\bibinfo {author} {\bibfnamefont {David~F.}\
  \bibnamefont {Wise}}, \bibinfo {author} {\bibfnamefont {John J.~L.}\
  \bibnamefont {Morton}}, \ and\ \bibinfo {author} {\bibfnamefont {Siddharth}\
  \bibnamefont {Dhomkar}},\ }\href@noop {} {\enquote {\bibinfo {title} {Using
  deep learning to understand and mitigate the qubit noise environment},}\ }
  (\bibinfo {year} {2020}),\ \Eprint {http://arxiv.org/abs/arXiv:2005.01144}
  {arXiv:2005.01144} \BibitemShut {NoStop}%
\bibitem [{\citenamefont {Cortes}\ and\ \citenamefont {Vapnik}(1995)}]{svm95}%
  \BibitemOpen
  \bibfield  {author} {\bibinfo {author} {\bibfnamefont {Corinna}\ \bibnamefont
  {Cortes}}\ and\ \bibinfo {author} {\bibfnamefont {V.~N.}\ \bibnamefont
  {Vapnik}},\ }\bibfield  {title} {\enquote {\bibinfo {title} {Support-vector
  networks},}\ }\href@noop {} {\bibfield  {journal} {\bibinfo  {journal}
  {Machine Learning}\ }\textbf {\bibinfo {volume} {20}},\ \bibinfo {pages}
  {273--297} (\bibinfo {year} {1995})}\BibitemShut {NoStop}%
\bibitem [{\citenamefont {Drucker}\ \emph
  {et~al.}(1996{\natexlab{b}})\citenamefont {Drucker}, \citenamefont {Burges},
  \citenamefont {Kaufman}, \citenamefont {Smola},\ and\ \citenamefont
  {Vapnik}}]{Drucker}%
  \BibitemOpen
  \bibfield  {author} {\bibinfo {author} {\bibfnamefont {H.}~\bibnamefont
  {Drucker}}, \bibinfo {author} {\bibfnamefont {C.~C.}\ \bibnamefont {Burges}},
  \bibinfo {author} {\bibfnamefont {L.}~\bibnamefont {Kaufman}}, \bibinfo
  {author} {\bibfnamefont {A.~J.}\ \bibnamefont {Smola}}, \ and\ \bibinfo
  {author} {\bibfnamefont {V.~N.}\ \bibnamefont {Vapnik}},\ }\bibfield  {title}
  {\enquote {\bibinfo {title} {Machine learning: Trends, perspectives, and
  prospects},}\ }\href@noop {} {\bibfield  {journal} {\bibinfo  {journal}
  {Advances in Neural Information Processing Systems}\ }\textbf {\bibinfo
  {volume} {9}},\ \bibinfo {pages} {155--161} (\bibinfo {year}
  {1996}{\natexlab{b}})}\BibitemShut {NoStop}%
\bibitem [{\citenamefont {Chang}\ and\ \citenamefont {Lin}(2011)}]{libsvm}%
  \BibitemOpen
  \bibfield  {author} {\bibinfo {author} {\bibfnamefont {Chih-Chung}\
  \bibnamefont {Chang}}\ and\ \bibinfo {author} {\bibfnamefont {Chih-Jen}\
  \bibnamefont {Lin}},\ }\bibfield  {title} {\enquote {\bibinfo {title}
  {Libsvm: A library for support vector machines},}\ }\href {\doibase
  10.1145/1961189.1961199} {\bibfield  {journal} {\bibinfo  {journal} {ACM
  Trans. Intell. Syst. Technol.}\ }\textbf {\bibinfo {volume} {2}} (\bibinfo
  {year} {2011}),\ 10.1145/1961189.1961199}\BibitemShut {NoStop}%
\bibitem [{\citenamefont {Fan}\ \emph {et~al.}(2005)\citenamefont {Fan},
  \citenamefont {Chen},\ and\ \citenamefont {Lin}}]{Fan2005}%
  \BibitemOpen
  \bibfield  {author} {\bibinfo {author} {\bibfnamefont {Rong-En}\ \bibnamefont
  {Fan}}, \bibinfo {author} {\bibfnamefont {Pai-Hsuen}\ \bibnamefont {Chen}}, \
  and\ \bibinfo {author} {\bibfnamefont {Chih-Jen}\ \bibnamefont {Lin}},\
  }\bibfield  {title} {\enquote {\bibinfo {title} {Working set selection using
  second order information for training support vector machines},}\ }\href@noop
  {} {\bibfield  {journal} {\bibinfo  {journal} {J. Mach. Learn. Res.}\
  }\textbf {\bibinfo {volume} {6}},\ \bibinfo {pages} {1889–1918} (\bibinfo
  {year} {2005})}\BibitemShut {NoStop}%
\end{thebibliography}%

\end{document}